\documentclass[fleqn,usenatbib]{mnras}

\usepackage{newtxtext,newtxmath}

\usepackage[T1]{fontenc}

\DeclareRobustCommand{\DA}[3]{#2}
\let\DAthebibliography\thebibliography
\def\thebibliography{\DeclareRobustCommand{\DA}[3]{##3}\DAthebibliography}
\DeclareRobustCommand{\daname}[3]{#2}
\let\dathebibliography\thebibliography
\def\thebibliography{\DeclareRobustCommand{\daname}[3]{##3}\dathebibliography}

\usepackage{graphicx}	
\usepackage{amsmath}	

\newcommand{\bagpipes}{{\sc bagpipes}}
\newcommand{\simstack}{{\sc simstack}}

\title[Dust Temperature Evolution]{Exploring the Evolution of Dust Temperature using Spectral Energy Distribution Fitting in a Large Photometric Survey}

\author[G. T. Jones and E. R. Stanway]{
Gareth T. Jones,$^{1}$\thanks{E-mail: g.jones.6@warwick.ac.uk}
Elizabeth R. Stanway$^{1}$
\\
$^{1}$Department of Physics, University of Warwick, Gibbet Hill Road, Coventry CV4 7AL, UK
}

\date{Accepted XXX. Received YYY; in original form ZZZ}

\pubyear{2023}

\begin{document}
\label{firstpage}
\pagerange{\pageref{firstpage}--\pageref{lastpage}}
\maketitle

\begin{abstract}
Panchromatic analysis of galaxy spectral energy distributions, spanning from the ultraviolet to the far-infrared, probes not only the stellar population but also the properties of interstellar dust through its extinction and long-wavelength reemission. However little work has exploited the full power of such fitting to constrain the redshift evolution of dust temperature in galaxies.
To do so, we simultaneously fit ultraviolet, optical and infrared observations of stacked galaxy subsamples at a range of stellar masses and photometric redshifts at $0<z<5$, using an energy-balance formalism.
However, we find UV-emission beyond the Lyman limit in some photometric redshift selected galaxy subsamples, giving rise to the possibility of contaminated observations.
We carefully define a robust, clean subsample which extends to no further than $z\sim2$.
This has consistently lower derived temperatures by $4.0^{+5.0}_{-1.9}$\,K, relative to the full sample.
We find a linear increase in dust temperature with redshift, with $T_d(z)=(4.8\pm1.5) \times z + (26.2\pm1.5)\ \mathrm{K}$.
Our inferred temperature evolution is consistent with a modest rise in dust temperature with redshift, but inconsistent with some previous analyses.
We also find a majority of photometrically-selected subsamples at $z>4.5$ under-predict the IR emission while giving reasonable fits to the UV-optical.
This could be due to a spatial disconnect in the locations of the UV and IR emission peaks, suggesting that an energy-balance formalism may not always be applicable in the distant Universe.
\end{abstract}



\section{Introduction}
Understanding its dust content is crucial to inferring accurate information about a galaxy.
Dust plays an important role in  physical processes on both galactic and sub-galactic scales.
It absorbs incident stellar radiation, scattering and thermally re-radiating the energy at infrared (IR) wavelengths \citep[e.g.][]{1996A&A...306...61B, 1999ApJ...521...64M, 2001ApJ...548..296W}.
This strongly reshapes the spectral energy distribution (SED) emitted by a galaxy. Thus, when extracting galaxy properties from SEDs it is crucial to consider the dust content of the galaxy, dust extinction laws and dust reemission models \citep[e.g.][]{2000ApJ...533..682C, 2008MNRAS.388.1595D, 2019A&A...622A.103B}.

This is particularly true at high redshifts, where dust properties such as grain size, temperature and composition are unknown, which hampers the inference of other galaxy parameters \citep[e.g.][]{2020MNRAS.497..956S,2022ApJ...939L..27C}.
The mean dust blackbody temperature in a galaxy ($T_d$) is one such parameter. Since dust is irradiated by stars, and generated by processes within stars or at their termination in supernovae, accurately constraining the dust masses in distant galaxies can provide important constraints on their stellar populations. Dust heated to higher temperatures may form differently sized grain distributions and potentially have different compositions to that with less thermal processing. The resultant different extinction and emission curves can result in misinterpretation of star-formation rate, galaxy colour and potentially both stellar and dust masses \citep[e.g.][]{2013ApJ...775L..16K,2015ApJ...806..259R,2020ARA&A..58..529S}.

The dust also mediates the next generation of star formation in a system, providing sites where hydrogen gas can be triggered into collapse and star formation \citep[e.g.][]{2004ApJ...604..222C,2018MNRAS.474.1545C,2022MNRAS.514.1461R}. However, dust radiates and thus cools extremely efficiently \citep[][]{1986A&A...160..295D,2022HabT.........1G}, and so when observed in emission must be close to an equilibrium with irradiation from its environment, or, at high redshift, with the thermal component of the cosmic microwave background radiation (which scales in temperature as $2.7\times(1+z)$\,K). 

Thus an evolution in dust temperature with redshift is expected if the irradiating population is younger, more luminous in the far-ultraviolet, differs in metallicity or otherwise changes with cosmic time. A consensus on the dust temperature evolution with redshift ($T_d-z$) is yet to be reached.
Observations of high-$z$ ($z>6$) galaxies have derived temperatures ranging from extremely low, close to the rest-frame Cosmic Microwave Background (CMB) temperature \citep[$T_d<30$\,K,][]{2022MNRAS.515.1751W}, up to very high outliers \citep[$T_d>80$\,K,][]{2018MNRAS.477..552B,2020MNRAS.493.4294B}.
This allows for many expected $T_d-z$ relations to be inferred, from negligible dust temperature evolution \citep[e.g.][]{2019MNRAS.489.1397L,2022ApJ...930..142D}, to linear \citep[e.g.][]{2018A&A...609A..30S,2020ApJ...902..112B} or even near exponential increases \citep[e.g.][]{2022MNRAS.516L..30V}.

The large uncertainty in evolution can be traced back to the combination of two factors: the data sample and the methodology used. The thermal emission must be observed primarily at wavelengths which lie redwards of the atmospheric cut-off in the mid- to far-infrared.
Observing in the infrared thus requires space telescopes. These are necessarily limited in size, and so have large point spread functions (PSFs). Combined with the faintness of most sources in the infrared, this leads to difficulties with interlopers, blending and reliably characterising objects.
This results in limited IR data for individual high-$z$ galaxies. Hence, dust properties are often derived from $<4$ data points, none of which may straddle the peak of the dust emission \citep[e.g.][]{2017ApJ...837L..21L,2020MNRAS.493.4294B,2022MNRAS.515.1751W}.

Limited data means that most IR dust SED modelling assumes that the emission follows modified blackbody (grey-body) or power-law prescriptions with few free parameters \citep[e.g.][]{2020ApJ...902..112B,2022MNRAS.516L..30V}, rather than using detailed radiative transfer models or multiple dust components \citep{2000MNRAS.313..734E,2007A&A...461..445S,2008MNRAS.388.1595D}.
With any of these emission models, assumptions must be made about at least some of the dust properties, such as their geometry relative to the radiation source and composition, which ultimately can influence any derived dust parameters \citep[][]{2007ApJ...657..810D}.
In addition, most dust modelling and parameter inference is done presuming that the dust emission occurs in isolation.
Stellar population synthesis (SPS) models have been developed since the late 1960s \citep[][]{1968ApJ...151..547T} but simultaneous modelling of IR with UV-optical data has only been done more recently \citep[see][]{2013ARA&A..51..393C}.
The incident starlight is the main energy source for exciting dust emission, and thus the emission spectrum, as well as intensity, is affected by the properties of the starlight which is irradiating the dust \citep[hardness, total energy budget, etc.,][]{2021ApJ...917....3D}.
This connection between these two components needs to be accounted for to achieve accurate modelling \citep{2022MNRAS.514.5706J}.

When modelling panchromatic SEDs, most approaches use an energy-balance formalism, whereby dust-attenuated energy from stellar UV-optical emission is distributed across IR dust emission components, assuming 100 per cent efficiency \citep{1999A&A...350..381D,2008MNRAS.388.1595D}.
However, this makes the assumption that the main heating source of the observed dust is the peak stellar emission observed in the UV-optical.
Studies involving high spatial resolution observations have shown there to be spatial offsets between the peak emission in the rest-frame UV and far-IR observations of some galaxies, even in the absence of any evidence for active galactic nuclei (AGN) activity.
This has been observed at all redshifts, with even local galaxies having inhomogeneities in the distribution of the stellar and dust emission components \citep[e.g.][]{2015MNRAS.452...54M,2019ApJ...882..107R,2022MNRAS.515.3126I,2022A&A...665A.137S}.
Spatial separation in the peak of emission could suggest that the stellar population heating the dust is highly obscured and thus is different to that observed in the UV-optical data.

Such a disconnect fundamentally invalidates a key assumption in the energy-balance approach to SED fitting, and so could potentially have an impact on any derived physical parameters.
However this may not necessarily be the case. In studying distant galaxies, particularly in the infrared, we are often constrained to consider the unresolved light generated from many individual star forming regions, old stellar populations, and their associated dust clouds. While the energy-balance formalism may break on small scales, some observed global scaling relations hint that it holds on average across many such regions. This is seen, for example, in the strong relationship observed between the UV colour (and hence inferred dust extinction) and the infrared emission (and hence inferred dust reprocessing) in normal star-forming galaxies \citep[e.g.][]{1999ApJ...521...64M,2014ApJ...796...95C,2020ARA&A..58..529S}.
Recent work has also shown that inhomogeneites in the emission components in resolved local galaxies do not affect the SED-fitting performance \citep{2022A&A...665A.137S}. A detailed resolved study was used to model the effects of increasing distance and redshift, and so as a proxy to predict the accuracy of SED modelling techniques in high-$z$ dusty galaxies with very complex geometry.
However, little work has explored simultaneous fitting of UV, optical (i.e. stellar) and IR (i.e. dust) photometric data in higher-$z$ galaxies directly.

The paper presents a panchromatic analysis of SED-inferred dust temperature evolution and is structured as follows. In Section~\ref{sec:methods}, we add motivation for this work before presenting the sample used for this analysis, the selection of observational data for these galaxies, our stacking methodology for the IR data, and our modelling and fitting methodology.
Section~\ref{sec:full_results} presents the results from fitting the COSMOS2020 photometric redshift-derived galaxy sample, including derived $T_d-z$ relations. We then consider contamination issues within the sample and develop a methodology to remove contamination. We generate a golden sample and consider its properties. Section~\ref{sec:discussion} discusses our derived $T_d-z$ relations, placing them in context of analysis done by other authors, and considering their implications.
We present a summary of our main findings in Section~\ref{sec:conclusions}. Except where otherwise specified, we adopt a standard $\Lambda$CDM cosmological model in which $\Omega_\Lambda=0.7$, $\Omega_{\mathrm{M}}=0.3$, and $H_0=70$\,$\mathrm{km s^{-1} Mpc^{-1}}$.

\section{Data and Methodology} \label{sec:methods}

\subsection{Strategy and Motivation}

In this work, we aim to investigate the $T_d-z$ relation out to $z\sim10$ and consider whether any information about the dusty component of these galaxies can be inferred (i.e. spatial separations between the UV-optical and IR emission) using the SED fitting technique. Unlike previous work in this area, we will simultaneously fit both the stellar and dust components, accounting for the well-established formalism of global energy-balance in each SED, to derive dust temperatures on a large, deep and well-studied photometric data set. 

Energy balance approaches \citep[e.g.][]{1999A&A...350..381D,2008MNRAS.388.1595D} assume that ultraviolet photons generated from young stellar populations are absorbed by dust grains, heating and modifying them. This energy is then reemitted as thermal radiation, described crudely by a simple blackbody, but with an emission spectrum modified by the dust grain size distribution and composition. This modified blackbody is known as a grey-body. Analytic and numerical models have been proposed for these grey-bodies \citep[e.g.][]{2000ApJ...539..718C, 2007ApJ...657..810D}. One key component of these analytic formalisms is the shape (i.e. the hardness) of the ultraviolet spectrum causing the heating. Hard UV photons can photo-evaporate small dust grains or sublimate chemical species from their surfaces, while high energy photons can shatter grains.  In each case, the dust emission curve is modified. As \citet{2021ApJ...917....3D} demonstrated, the use of different stellar population templates as the irradiating source leads to a difference in temperature profile and distribution of grain properties in model dust clouds.
Observations of infrared-luminous galaxies have shown that some classes of objects (e.g. submillimetre galaxies) have higher derived dust temperatures than those observed in other classes (such as LIRGs).
These properties correlate with many aspects of the underlying stellar population, particularly the inferred star formation rate \citep[e.g.][]{2010MNRAS.409...75H, 2012MNRAS.425.3094C}. Further, \citet{2022MNRAS.514.5706J} demonstrated that different dust temperatures, luminosities and masses could be derived from the same far-infrared photometry, if the presumed irradiating spectrum template is changed. This is a particular problem given the difficulty of observing galaxies in the far-infrared, and the low signal-to-noise or sparse photometric data points that can be obtained for any given system. Constraints on the shape of the curve are often weak, with measurements of the blackbody peak (i.e. the dust temperature) somewhat degenerate with the wavelength-dependent emissivity profile (i.e. the grain properties).

Since the stellar population is intimately associated with the shape, temperature and strength of the dust emission curve, simultaneous, self-consistent fitting of the observed spectral energy distributions of dust-attenuated starlight and thermal emission should provide an improved constraint on the dust temperature, if the energy-balance formalism holds.  By contrast, if the dust is optically-thick, only dust emission will be observed (the optical light is highly absorbed by definition) and so the dust emission curve expected from the observed starlight will not provide a good match to the data. Here we aim to observe optically-selected galaxies, and hence are in a regime where the energy-balance formalism has been widely applied in past work \citep[e.g.][]{2008MNRAS.388.1595D,2018MNRAS.480.4379C,2019A&A...622A.103B}. However we also aim to consider whether cases exist in which it is clearly at fault.

To explore the use of UV-FIR full spectrum SED fitting, we will exploit data from a large, deep, wide-field galaxy survey with good photometric coverage.
The COSMOS2020 catalogue \citep[][]{2007ApJS..172....1S,2022ApJS..258...11W} contains deep UV-optical observations of galaxies with photometric redshifts extending to $z\sim10$, the positions of which can be combined with IR/submillimetre maps available through the Herschel Extragalactic Legacy Project \citep[HELP,][]{2010A&A...518L...2P,2012MNRAS.424.1614O} and SCUBA-2 Cosmology Legacy Survey \citep[S2CLS,][]{2017MNRAS.465.1789G}. 

While IR detections are not frequently observed for individual galaxies, the mean properties of samples of galaxies can be examined through stacking the object locations in infrared maps.
This is scientifically useful since galaxies with similar physical properties taken from optical observations should have, on average, similar IR properties \citep{2018A&A...609A..30S, 2022MNRAS.516L..30V}.

\subsection{Sample Selection}
In order to evaluate the evolution of dust temperature we draw observational data from the Cosmic Evolution Survey (COSMOS) observations \citep{2007ApJS..172....1S}, compiled into the COSMOS2020 catalogue \citep{2022ApJS..258...11W}. We select this survey as it combines deep-field sensitivity with a large area (allowing bright galaxies to be studied) and extensive multiwavelength photometric coverage of galaxy SEDs that extend from the far-ultraviolet to submillimetre wavelengths.
The recent COSMOS2020 catalogue compiles together all recent observations taken of the field and detects faint sources with a magnitude of $K_S=25.2$ (AB), an improvement of nearly one magnitude over previously released versions. This extra depth increases the number of photometric galaxy candidates available to stack at all redshifts, allowing such analyses to probe typical galaxies and also permitting accurate flux measurements when stacking.

The COSMOS team have released two COSMOS2020 photometry catalogues in which the flux is measured in different ways.
We focus on the catalogue using the newly developed FARMER/{\sc{LePhare}} profile-fitting tool which is marginally superior at detecting faint sources over the CLASSIC catalogue.
The FARMER catalogue reports 868\,048 galaxies out to an estimated photometric redshift of $\sim$10.
We focus on star-forming galaxies. We follow the prescription adopted by \citet{2022MNRAS.516L..30V} to split the sample into star-forming and quiescent galaxies based upon the $NUV-r$ versus $R-J$ selection, where galaxies with $M_{\mathrm{NUV}}-M_{r}>3(M_{r}-M_{J})+1$ and $M_{\mathrm{NUV}}-M_{r}>3.1$ are considered quiescent \citep{2013A&A...556A..55I}.
We also focus on galaxies with masses in the range log(M/M$_\odot$)$=9.5-12.0$ to ensure reliable detection at all redshifts and to avoid the extreme cosmic variance that affects rare very massive galaxies.
This reduces the sample under consideration to 111\,277 star-forming galaxies.

 We bin subsamples of galaxies by mass and photometric redshift to increase the signal-to-noise ratio (S/N) on the typical spectral energy distribution, especially in the infrared regime where data is limited. This averages over a range of star formation histories and evolution pathways, and allows us to look for trends in typical galaxy properties with mass or redshift. We separate the galaxies into 4 mass bins ($\mathrm{log(M/M}_{\odot})$=9.5-10.0, 10.0-10.5, 10.5-11.0, and 11.0-12.0), and 23 photometric redshift bins (spaced $\Delta z=0.25$ between $z=0$ and $5$, $\Delta z=0.5$ between $z=5$ and $6$, and one bin encompassing $z=6-10$).

\subsection{UV, Optical and Near-IR Photometric Data}
The COSMOS2020 catalogue contains data for individual galaxies in the UV, optical and near-IR (NIR) wavelengths.
Included data is sourced from GALEX, CFHT/MegaCam, HST/ACS, Subaru/HSC, Subaru/Suprime-Cam, VISTA/Vircam and Spitzer/IRAC, covering a spectral range from 1340\,\AA\ to 9.6\,$\mu$m in the observed frame.
Observations in each photometric band are combined together to determine the flux for a typical galaxy within each sample.
Within the catalogue, the amount of Galactic extinction in each object's line of sight is calculated using the \citet{1998ApJ...500..525S} dust extinction map values.
We use the \citet[][]{1999PASP..111...63F} interstellar attenuation law, with the Milky Way $R_V=3.07$, to correct for Galactic extinction in all the photometric measurements.

\subsection{Mid-to-Far-IR Data}
While catalogued fluxes are stacked in the UV and optical, this approach is not appropriate for the infrared, where the majority of sources are undetected. Instead stacking must occur in the image plane, where the reduced noise in a stacked image may permit the joint detection of sources that would individually fall below the selection limit.

To extract infrared data for each subset of galaxies, we follow the same stacking methodology as \citet{2022MNRAS.516L..30V}, apart from the fact that we have a finer redshift grid spacing.
This method takes as its source archival infrared maps and the locations of objects from the COSMOS2020 catalogue.
The maps used are the 24\,$\mu$m Spitzer/MIPS Spitzer Enhanced Imaging Product \citep[SEIP;][]{2010A&A...512A..78B}, the 100 and 160\,$\mu$m Herschel/PACS maps from PACS Evolutionary Probe \citep[PEP;][]{2010A&A...518L...2P}, the 250, 350 and 500\,$\mu$m Herschel/SPIRE maps from the Herschel Multi-tiered Extragalactic Survey \citep[HerMES;][]{2012MNRAS.424.1614O}, and the 850\,$\mu$m SCUBA map from S2CLS \citep{2017MNRAS.465.1789G}. These were obtained from the respective repositories and compiled into a single data package associated with the \simstack\ software \citep[for further details see][]{2022MNRAS.516L..30V}. 

To permit direct comparison with previous work that analysed only the far-infared data on the same objects, we use the most recent mid-2022 release of the \simstack\ algorithm \citep{2013ApJ...779...32V, 2022MNRAS.516L..30V}\footnote{The \simstack\ software  adopts the Planck18 cosmology, with $\Omega_\Lambda=0.685$, $\Omega_{\mathrm{M}}=0.313$, and $H_0=67.4$\,$\mathrm{km s^{-1} Mpc^{-1}}$.}. 
The \simstack\ algorithm generates a model point spread function (PSF) for each photometric map. This PSF model is scaled against the image at the 2D positions in a given catalogue, and in narrow slices of photometric redshift, to generate a mock image and compare its covariance with the initial image. The model which provides the best-fits is interpreted as the average flux density for the objects in the bin. Thus a single flux measurement per band is generated for each bin or input catalogue.
By using the correlation between objects in a catalogue, \simstack\ has the potential benefit that it overcomes inherent biases present when stacking images dominated by confusion noise \citep{2022MNRAS.516L..30V}. 

\simstack\ was originally developed as part of a project to evaluate the contributions of galaxies to the cosmic infrared background, and further modified to consider dust temperature evolution over cosmic time \citep{2013ApJ...779...32V, 2022MNRAS.516L..30V}. Its outputs include a dust temperature estimate derived by fitting the infrared points in isolation from the UV-optical stellar population. It uses a simple grey-body spectral curve, truncated by a power law at low wavelengths. No consideration was made of ultraviolet, optical or near-IR data, except as a source of galaxy coordinates, and no spectral fitting of the stellar population undertaken. 

We restrict the analysis to keep only subsamples which have detections equal to or above 3-sigma in at least four of the IR bands in the stacked galaxy data, reducing the number of subsamples from 92 to 73.
We note that this should preferentially remove subsamples dominated by low temperature sources, since they would be fainter in the long wavelength bands, although it will also impact the highest redshift subsamples since these are close to the detection limit. For any other subsample where there was still a non-detection (i.e. negative flux), a 3-sigma upper limit based upon the \simstack-derived flux uncertainty was used instead. This step ensures sufficient datapoints to constrain free parameters in the models.

\subsection{SED Fitting with \bagpipes} \label{sec:bagpipes}

A simultaneous analysis of all available UV, optical and infrared flux data, stacked for each galaxy luminosity-redshift subsample, was undertaken using the Bayesian Analysis of Galaxies for Physical Inference and Parameter EStimation fitting algorithm \citep[\bagpipes,][]{2018MNRAS.480.4379C}.
This Bayesian spectral fitting code utilises a chosen stellar population synthesis model in combination with nebular emission libraries, a dust attenuation prescription, and a multi-component dust emission model, to fit and interpret the integrated light from a galaxy. We use the modified version of \bagpipes\ reported in \citet{2022MNRAS.514.5706J}, which permits a broader range of SPS and dust emission models to be used than the original version of \citet{2018MNRAS.480.4379C}.

In this analysis, we use the Binary Population and Spectral Synthesis (BPASS) v2.2.1 models \citep{2017PASA...34...58E,2018MNRAS.479...75S} to describe the stellar population, with an empirical, multi-temperature emission component based on the \citet{2008MNRAS.388.1595D} formalism to describe the dust emission.

The BPASS models are a set of SPS models which incorporate binary evolution in addition to single star evolutionary pathways.
The binary models are full, detailed stellar evolution models, in which the interior structure of the star is calculated, but allow for mass-loss or gain through binary interactions.
Nebular emission for the models is calculated using the {\em Cloudy} radiative transfer code \citep{2017RMxAA..53..385F}.
This illuminates a nebular cloud with the BPASS spectra, where the cloud is assumed to have an electron density of 200 atoms cm$^{-3}$, a fixed nebular ionization parameter, log $U_\mathrm{neb} = -3.0$, and a spherical geometry. Since we do not expect nebular emission to significantly alter the interpretation of the broadband stellar photometry, we do not vary the nebular emission library.

Dust emission is incorporated using the \citet{2008MNRAS.388.1595D} model algorithm. This is a set of empirical, physically motivated models for the thermal emission from radiation-heated dust grains.
There are two models: one for the dense birth cloud around young stars which dissipates on the order of a few Myrs \citep[e.g.][]{2015MNRAS.449.1106H,2017A&A...601A.146C,2020SSRv..216...50C}, and another for the cooler dust found within the ISM.
These are each made up of multiple components: a PAH emission template (for which we use the \citet{2021MNRAS.503.2598B} template), a mid-IR emission component (generated from equally weighted 130 and 250~K greybodies), a hot grain component and, in only in the ISM, a cold grain component (both modelled using a variable temperature grey-body).
This gives the model seven free parameters: five relative contribution of components ($\zeta_{PAH}^{BC}$, $\zeta_{PAH}^{ISM}$, $\zeta_{W}^{BC}$, $\zeta_{W}^{ISM}$, $\zeta_{C}^{ISM}$) and two dust temperatures ($T_W$, $T_C$).
More detailed information for both the stellar and dust emission models and their use can be found within \citet{2022MNRAS.514.5706J}.

SEDs were created from these models based upon two stellar populations: an old stellar population with an age of $>$0.1~Gyr, and a less massive young stellar population with fixed age of 5~Myr. The masses of both components were fitted for simultaneously.
The old population was modelled using a simple parametric, delayed-$\tau$ star formation history (SFH), which defines a star formation rate, SFR~$\propto te^{-t/\tau}$, where t is the time elapsed between the onset of star formation and the epoch of observation and $\tau$ is a parameter describing the exponential decay time-scale of the SFR.
The $\tau$ parameter is fitted for in the old population.

A selection of metallicities between $Z=0.001$ and $Z=0.020$ were  trialled.
There were no clear trends with metallicity, and no strong preference for one metallicity over another in the case of most bins. This is consistent with previous work on SED fitting which has shown little discriminating power between metallicities in most stellar populations \citep[see e.g.][]{2013ARA&A..51..393C}. Thus the metallicity presumed for both populations in the following analysis was fixed at $70\%$ solar (where $\mathrm{Z_{\odot}}=0.020$).

We use a single dust attenuation law \citep{2000ApJ...533..682C} and birth cloud attenuation multiplicative factor (a parameter in \bagpipes\ which accounts for the increased attenuation in very young star forming regions) set as $\eta=2$, with the birth cloud dispersion age set as 5~Myr.
The attenuation in the V-band, $A_V$, was allowed to vary.
For the dust emission models, both the relative contributions of components and dust temperatures were fitted for, where the cold grain temperature was fitted in the range $14-55$~K and the warm grain temperature in range $25-155$~K, using flat priors.

We note that an additional heating source comes from CMB photons. At high redshift, these can approach temperatures of $\sim$20\,K, comparable to the minimum temperature of stellar-heated dust clouds. The heating becomes non-negligible around $z\sim4$ \citep[][]{2013ApJ...766...13D}. Since the majority of our results are obtained well below this redshift, we do not implement CMB heating of dust in our model.

\begin{figure*}
    \includegraphics[width=\textwidth]{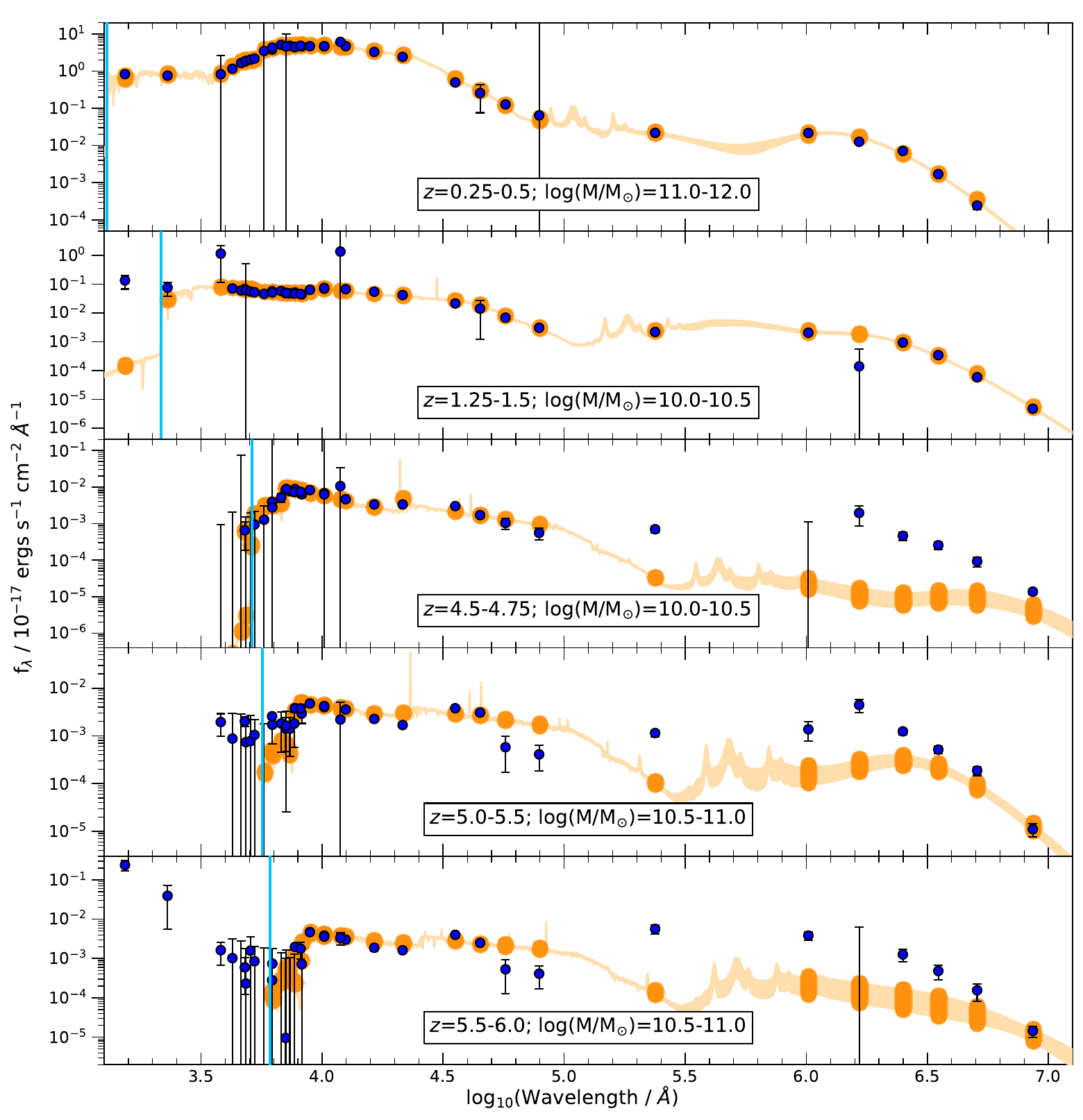}
    \caption{Best-fit model from \bagpipes\ for a selection of subsamples. The sample's photometric redshift and mass bin is labelled on the bottom of each plot. The best-fit model spectrum is shown in orange, with the orange dots representing the model flux in each filter, where the width indicates the uncertainty within the model. The observational data for each filter is overlaid as blue circles. Each model has been redshifted to the mean redshift of the bin ($z=0.42$, 1.37, 4.64, 5.22, 5.70 from top to bottom). The solid light blue line indicates the Lyman limit for that sample (at 1295, 2161, 5144, 5673, 6110\,\AA\ from top to bottom). Some of the samples still have detections bluewards of their Lyman limit, which is only likely if there was contamination within the sample.}
    \label{fig:model_grid_5}
\end{figure*}

The top panel in Fig.~\ref{fig:model_grid_5} shows an example best-fit model (orange spectrum) to the observational data (blue circles) from the stacked galaxies with $\mathrm{log(M}/\mathrm{M_{\odot}})=11.0-12.0$ at $z=0.25-0.5$, where almost all data points have modelled fluxes within their observational uncertainty. As Fig.~\ref{fig:model_grid_5} illustrates there are more optical data points available than infrared. In total there are 2 UV, 25 optical, 4 near-IR and 7 far-IR filters. This means that the optical data dominates the fits, and thus the stellar component is well constrained. The dust emission component will only be constrained by the 7 far-IR data points and possibly a couple of the near-IR, depending on redshift. The amount of flux available for dust emission comes from extincted blue, ultraviolet radiation according to the energy balance formalism used. As the dust emission component is fit simultaneously, it could, in principle affect the stellar population fit by requiring a higher photon production rate. However, the imbalance in data coverage means that this is unlikely to happen. The extincted UV-optical is far more tightly constrained, with the infrared emission component having little effect due to the relative weighting of the photometric points. 

The dust temperature, i.e. the peak of the dust emission, will be constrained by only 2-3 points. In certain cases, due to either low numbers of galaxies in a given subsample or them being too faint to be detected collectively in the IR stack, the points around the dust emission peak may be non-detections and give unrealistic dust temperatures. Thus, manual vetting of all model fits was done to ensure that the derived dust temperatures were realistic.

\section{Results of SED Fitting} \label{sec:full_results}

\subsection{Dust Temperature Evolution} \label{sec:temp_results}
\begin{figure}
    \includegraphics[width=\columnwidth]{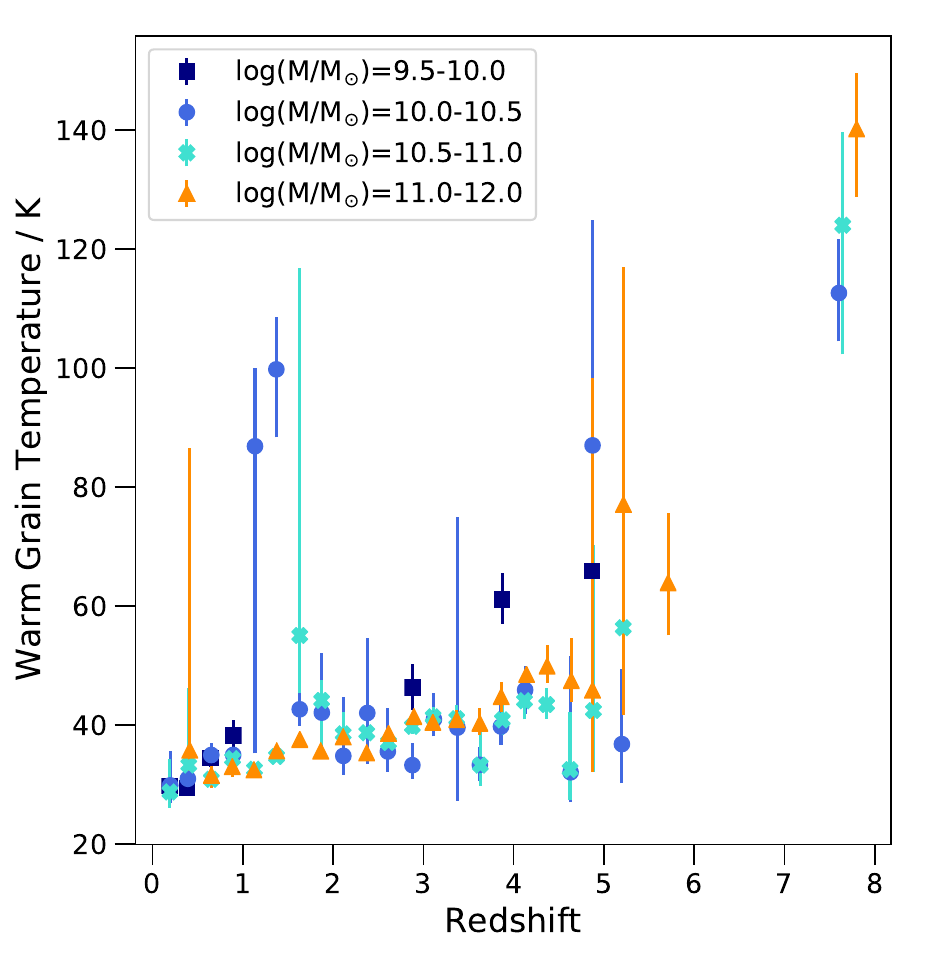}
    \caption{Derived warm grain temperature for each stacked galaxy sample. Different symbols indicate the different mass bins, with $\mathrm{log}(\mathrm{M}/\mathrm{M_{\odot}})=9.5-10.0$, $10.0-10.5$, $10.5-11.0$ and $11.0-12.0$ shown as dark blue squares, medium blue circles, light blue crosses and orange triangles, respectively.}
    \label{fig:temp_inc_bins}
\end{figure}

The \citet[][]{2008MNRAS.388.1595D} dust emission model has four components. The warm grain component of the model straddles the peak of the blackbody emission and thus its temperature can be considered the average dust temperature within a galaxy.
The derived warm grain temperatures from \bagpipes\ for each sample is shown in Fig.~\ref{fig:temp_inc_bins}.
The different mass bins are shown with different colours and marker symbols.
The distribution of temperatures appears to be bimodal: galaxy samples with low grain temperatures (<\,50\,K) with small uncertainties, while the others have higher grain temperatures (>\,60\,K) and/or larger uncertainty ranges.
These latter subsamples can be separated into two groups: those without the crucial 160~$\mu$m flux points to constrain the peak temperature, and those in which the IR model simply does not reproduce the flux seen in the observations.
Examples of these can be seen within the second and bottom three panels of Fig.~\ref{fig:model_grid_5}, respectively.

The second panel of Fig.~\ref{fig:model_grid_5} shows the SED of the galaxy subsample  with $\mathrm{log(M}/\mathrm{M_{\odot}})=10.0-10.5$ at $z=1.25-1.5$. This yields an outlying data point with a dust temperature of $100^{+11}_{-9}$\,K.
The best-fitting IR model reproduces most of the IR data points apart from the 160~$\mu$m band. This data point has a much larger uncertainty and a lower flux compared to photometry in its neighbouring filters.
The 160~$\mu$m filter lies around the peak of the infrared flux curve which can make it critical to get accurate dust temperature measurements. In its absence, \bagpipes\ uses the other 6 IR data points to generate its best-fitting model. However this is poorly constrained, allowing for a wide-range of permitted grey-body curves. 

Genuinely high dust temperatures in galaxies are a possibility and we do not exclude these from our analysis.
However it is clear than in the one subsample which is the largest outlier amongst our selection, the poor 160$\mu$m data may well be impacting the fit. Given this interpretation, it is inappropriate to retain this subsample in later analysis.

On Fig.~\ref{fig:temp_inc_bins}, we identify four redshift-luminosity galaxy subsamples at $0<z<4.5$ for which the derived temperature is effectively unconstrained, giving a 16-84th percentile range which allows any temperature between $\sim$25 and 70\,K or above. Inspection of the stacked data for these samples, identifies them as being additional cases in which data points are missing, or are clearly subject to large photometric uncertainty, around the peak of the dust curve. Since no strong constraint can be obtained for their dust temperatures, we omit these from further analysis of the dust temperature evolution, while noting that including their best-fit values would have negligible impact on subsequent results.

In the remaining subsamples at $z>4.5$, \bagpipes\ produced excellent fits to the UV-optical-NIR spectral range, but these implied a dust emission curve clearly inconsistent with the observed far-IR data. Examples are shown in the bottom three panels of Fig.~\ref{fig:model_grid_5}. As a result, the derived dust temperatures are physically unmotivated, and cannot be considered reliable. They are thus excluded from further analysis.
Possible reasons for the disconnect between the optical and far-IR fits are discussed in Section~\ref{sec:disconnect}.
 
\begin{figure}
    \includegraphics[width=\columnwidth]{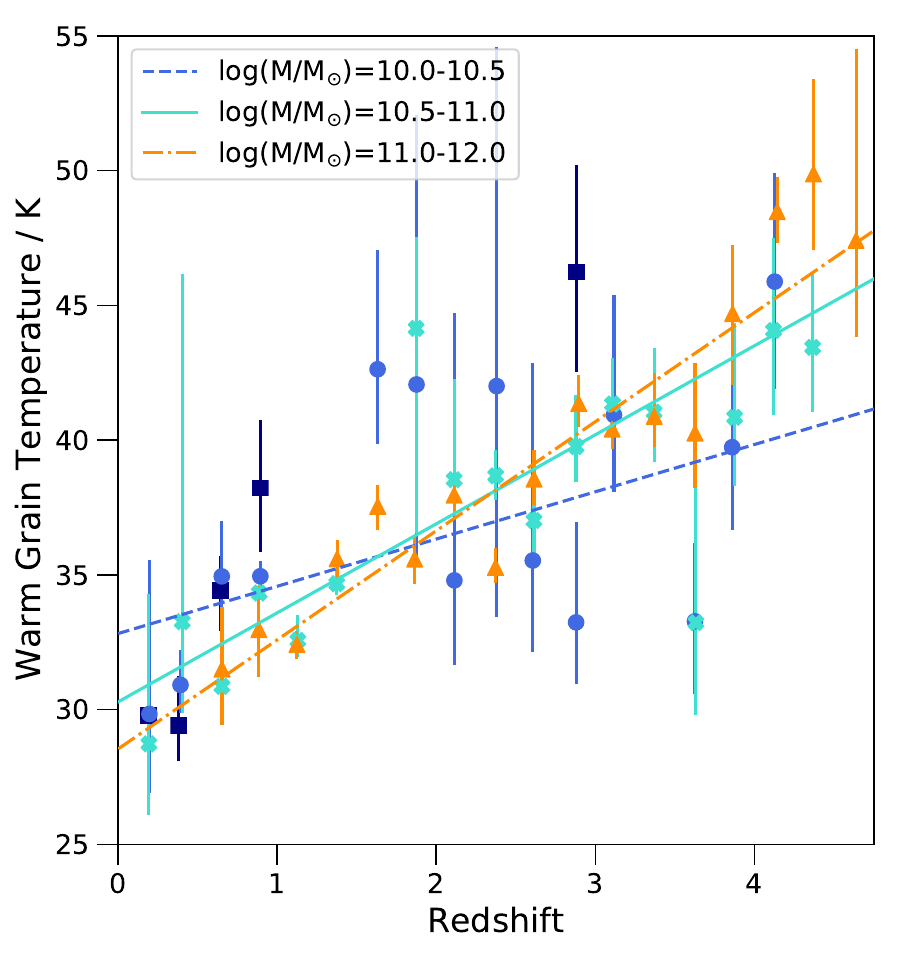}
    \caption{Same as Fig.~\ref{fig:temp_inc_bins} but after vetting the subsamples, removing those with missing 160~$\mu$m flux detections causing poor fits or subsamples where the IR model does not fit the observations. Included are linear least-square fits to individual mass bins, shown as a dashed medium blue, solid light blue, and dashed-dotted orange lines for the mass bins $\mathrm{log}(\mathrm{M}/\mathrm{M_{\odot}})=10.0-10.5$, $10.5-11.0$, and $11.0-12.0$, respectively. The lowest mass bin was not fitted due to a limited number of robust temperature measurements. The best-fitting parameters for these lines can be found within Table~\ref{tab:best_params}.}
    \label{fig:temp_inc_bins_vet_cut}
\end{figure}

Our derived dust temperature evolution with redshift is shown in Fig.~\ref{fig:temp_inc_bins_vet_cut}. Previous work \citep[e.g.][]{2018A&A...609A..30S} has suggested that temperature depends linearly on redshift. Following this prescription, we define the best fitting trend in temperature as 
\begin{equation}
T(z) = A z + B,
\label{eqn:linear}
\end{equation}
where $A$ and $B$ are constants, based on least squares minimisation for the trend in each mass range for which data permits.
The best-fitting parameters for the linear fits are presented in Table~\ref{tab:best_params}.
Note that when including the subsamples with poor constraints on the dust peak (and hence large temperature uncertainties), the values for $A$ and $B$ do not change significantly.

The cold grain temperature has a similar positive trend to the hot grain at lower derived temperatures, however it is not as well constrained. Arguably only the two longest wavelength data points have large contributions from these grains, which sit on the grey-body tail of the emission curve, and as a result,  there is not enough information to consistently constrain this parameter alongside the others.

\begin{table}
	\centering
	\caption{Best fitting parameters for the temperature dependence on redshift shown in Fig.~\ref{fig:temp_inc_bins_vet_cut}. Parameters $A$ and $B$ are defined by Equation \ref{eqn:linear}.}
	\label{tab:best_params}
	\begin{tabular}{lcc}
		\hline
		Mass Bin & A / K per unit $z$  & B / K\\
		\hline
		$10.0-10.5$ & $1.8\pm0.7$ & $32.8\pm1.0$ \\
		$10.5-11.0$ & $3.3\pm0.5$ & $30.3\pm0.6$ \\
		$11.0-12.0$ & $4.0\pm0.5$ & $28.5\pm1.1$ \\
		\hline
	\end{tabular}
\end{table}

\subsubsection{Comparison to \simstack\ derived temperatures} \label{sec:simstack_temps}
An additional inbuilt feature within the \simstack\ software is an independent SED fitting analysis of the stacked IR data measurements.
This is detailed in \citet{2022MNRAS.516L..30V} and we give a brief outline here.
IR SEDs from stacked flux densities are fit using a hybrid blackbody model, whereby the mid-IR Wien side of the model is substituted with a power law of fixed slope, $\alpha=2.0$, and emissivity index, $\beta=1.8$ \citep[e.g.][]{2012MNRAS.425.3094C}.
This leaves the model with two free parameters: amplitude and observed dust temperature.
This simple model allows for fitting to any IR SED, including those with very few data points, but ignores the stellar component and shorter-wavelength dust components (i.e. PAH emission).

\begin{figure}
    \includegraphics[width=\columnwidth]{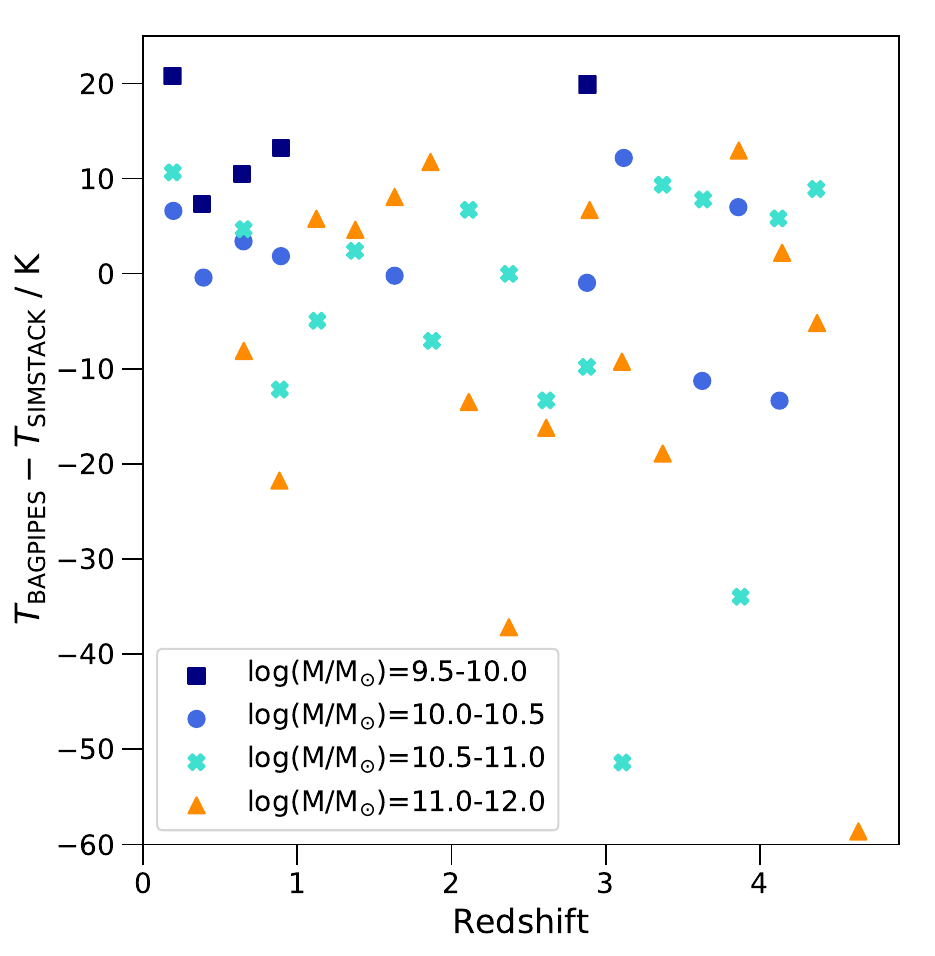}
    \caption{Comparison of the derived temperatures from \bagpipes\ and \simstack\ for all galaxies included in the vetted full sample, calculated as the \bagpipes\ derived value minus the \simstack\ value. The different symbols represent the different mass bins, with $\mathrm{log}(\mathrm{M}/\mathrm{M_{\odot}})=9.5-10.0$, $10.0-10.5$, $10.5-11.0$ and $11.0-12.0$ shown as dark blue squares, medium blue circles, light blue crosses and orange triangles, respectively. Errors have not been included in this plot for clarity.}
    \label{fig:comp_cut}
\end{figure}

The \simstack\ software derives a dust temperature for all subsamples using this simple model.
We compare these to the derived temperatures from our analysis in Fig.~\ref{fig:comp_cut}.
For the majority of the subsamples, the two values agree within $\sim20$\,K of each other.
In addition there are a handful of outliers at $2<z<5$ where \simstack\ estimates much higher temperatures ($35-60$\,K higher) than the \bagpipes\ fits. 
In total, 66\% of the subsamples lie within a 12\,K difference, and 87\% lie within a 20\,K difference.

We note that a large spread of $20$\,K, when the \bagpipes-derived temperatures are only $\sim30-45$\,K is somewhat concerning, and does highlight how different models can derive vastly differing temperatures.
Thus, a good understanding of the model being fit and its assumptions is required to ensure that the temperature derived and its uncertainties can be properly quantified \citep[e.g.][also see Section~\ref{sec:limits} for discussion]{2022arXiv221201915P}.

\subsection{Contamination Issues}\label{sec:contamination}
While \bagpipes\ produces robust fits for the high photometric redshift subsamples plotted in Fig.~\ref{fig:temp_inc_bins_vet_cut}, this does not necessarily imply that those fits can be straightforwardly interpreted in a physical context. Several of the high-$z$ sample stacks exhibit UV and optical detections below the Lyman limit.
Examples of this are the subsamples $\mathrm{log(M}/\mathrm{M_{\odot}})=10.5-11.0$ at $z=5.0-5.5$ and $z=5.5-6.0$, as shown in the bottom two panels of Fig.~\ref{fig:model_grid_5}.
The mean redshift of these stacks are $z=5.22$ and $z=5.70$, at which the Lyman limit occurs at 5673 and 6110~\AA, respectively.
However, there are filters bluewards of this limit that have strong  detections, implying a Lyman continuum escape fraction of near unity and zero intergalactic medium absorption, which should not be possible if all the galaxies included in the stack were in the redshift range.
This means that there must be contaminating lower-redshift galaxies, coming from either photometric redshift errors or through confusion with interlopers along the line of sight.
There is always a balance to be struck between sample completeness and contamination. However when looking at extremely faint populations, in which a few bright interlopers can significantly bias the stacked properties, optimising low contamination is vital. This is the case in the far-infrared regime. However the infrared also makes exclusion of contaminants more challenging. The long wavelengths in the IR-bands result in large PSFs,
so that a number of different galaxies within the instrument PSF may be contributing flux to the detected source. Hence the impact of contaminants within the sample is likely to be larger than in the optical.

To mitigate the impact of contaminants in the sample, we calculate the probability of chance alignment, $\mathrm{P_{Chance}}$, to select a subset of galaxies which exist in regions with a low density of foreground objects. We calculate the local number density of objects (stars and galaxies) within a projected radius of $30''$ from each target galaxy in the sample.
We use this local density to predict the chance that an aperture placed at random in this field would, by coincidence, contain a contaminating galaxy.  This is determined by the product of the local projected number density with a minimum radius that defines the aperture. 
We choose to use an aperture size appropriate to the Spitzer/MIPS 24$\,\mu$m band, and identify the probability of a coincident object occurring within its PSF FWHM of $6''$. The MIPS  24$\,\mu$m band is the shortest wavelength infrared filter used by \simstack. It therefore produces the most liberal (i.e. lowest) probability that chance alignment is contaminating the measured flux. For most of the far-infrared bands, a much larger photometric aperture is used, however the probability of any given contaminant galaxy being a strong emitter at these wavelengths is lower. 
The product of the local number density of sources with the area of a 6$''$ radius aperture thus defines the chance alignment probability; the probability that the measured flux originates in part or in whole from another source along the line of sight.

In subsequent analysis we opt to prioritise low contamination over high completeness by restricting our sample to sources with $\mathrm{P_{Chance}}<0.1$ (i.e.  less than 1 in 10 remaining galaxies in a given galaxy subset is likely to still be an interloper). Since the contamination is, in almost every bin, likely to lie from sources at lower redshifts along the line of sight, the selection should be uncorrelated to the properties of the galaxies in the subsample themselves. Since the properties of the galaxies in a high redshift bin has no dependence on structure along the line of sight, this is effectively a random sampling and provides significantly reduced sample sizes but a more robust view of the properties of distant galaxies in a sample.

This procedure reduces the sample (including all redshifts and mass bins) to 2944 galaxies. This represents only a few percent of the original catalogue. On the other hand, those few percent have not been selected for any particular infrared properties. Their dust curves should be representative of the entire sample. By contrast, for the remainder of the sample, while the selected galaxy {\it may} dominate the infrared flux, this is by no means guaranteed.  If interlopers along the line of sight have substantially different properties (or redshifts) to the target galaxies, even a small contamination fraction in the infrared would substantially bias the interpretation. This cannot be easily addressed with crude estimates of the contamination fraction, since it is sensitive to the properties of the interlopers in each case, and these will be different for each galaxy subsample. Thus in this analysis we opt to prioritise robustness, over completeness. We return to this point at the end of the next section.

\subsection{Photometric Redshift Robustness}
In order to check the robustness of the photometric redshift estimate for each galaxy, we perform an independent photometric redshift analysis. We
fit the UV, optical and near-IR data (excluding the mid- and far-IR) individually for each galaxy in the sample using \bagpipes. This was done using a delayed-$\tau$ SFH with a single stellar population, allowing age, mass, $\tau$ SFH parameter, extinction and redshift to vary with flat priors, while metallicity was  fixed at 70\% solar.

\begin{figure}
    \includegraphics[width=\columnwidth]{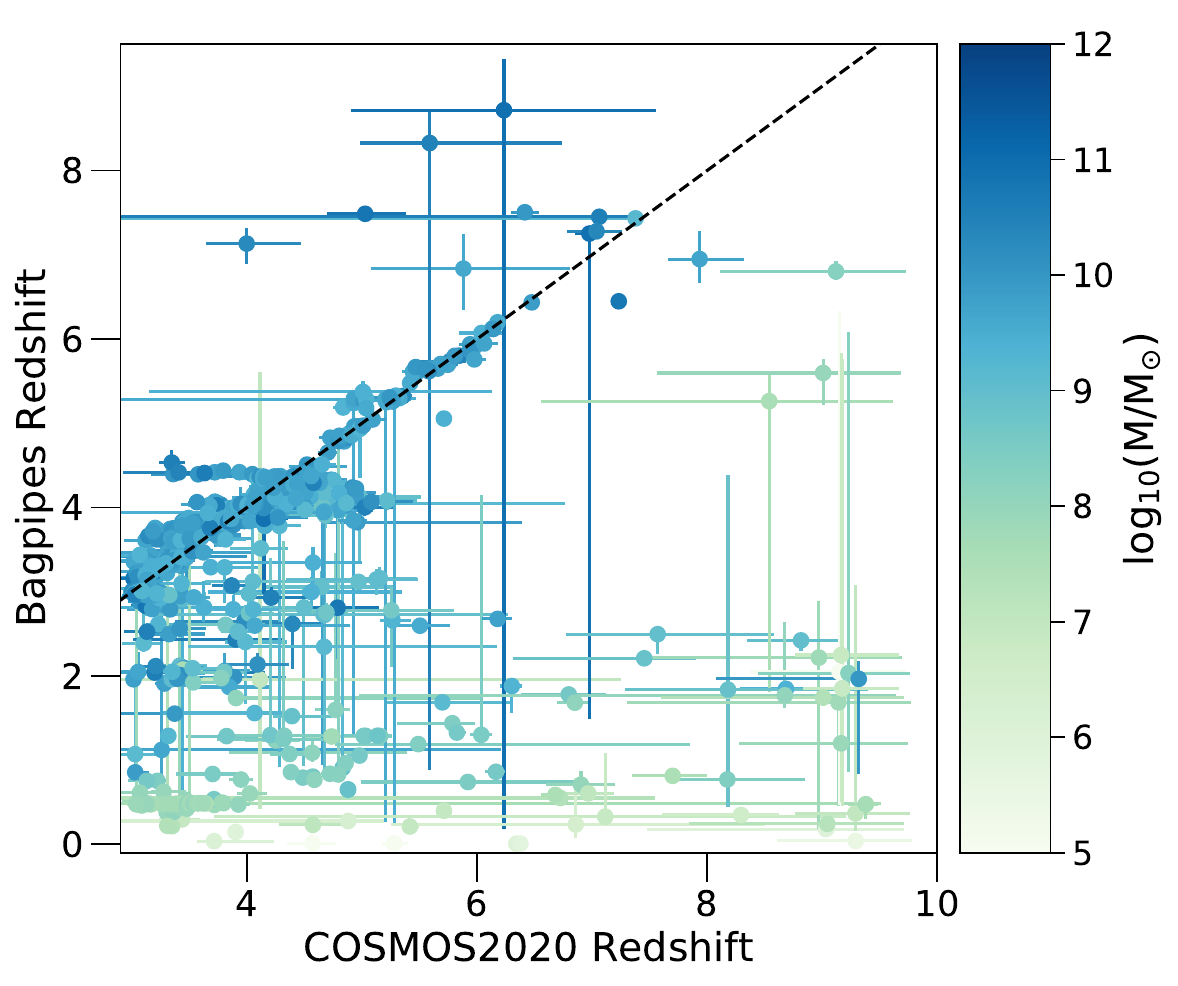}
    \caption{Derived redshift from fitting with \bagpipes\ compared to the derived photometric redshift from the COSMOS2020 catalogue for all galaxies remaining in the sample after a chance alignment probability cut. The black dashed line shows where the two derived values would be equal and the colourbar indicates the derived stellar mass from \bagpipes. All galaxies in the sample were fitted but this plot only shows those with COSMOS2020 photometric redshifts of $z>3$.}
    \label{fig:z_comp}
\end{figure}

The resulting values are compared to the COSMOS2020 catalogue values in Fig~\ref{fig:z_comp}. Fitting was performed on all galaxies in the sample, but the figure only shows those with COSMOS2020 redshifts of $z>3$ for clarity. For the majority of galaxies the derived \bagpipes\ redshift agrees with the COSMOS2020 value. However, there are clear and numerous catastrophic disagreements between the two estimates, for which a high redshift solution is favoured in one analysis, and a low redshift is favoured by the other. Significantly, these catastrophic disagreements are not symmetrically distributed.  Many of the galaxies assigned high photometric redshifts in the COSMOS2020 analysis are assigned a lower photometric redshift solution using \bagpipes. Given the intrinsic selection biases in any galaxy sample, and the decreasing typical fluxes of galaxies with increasing redshift, any low redshift solution that is comparable in probability to a high redshift solution should be formally favoured based on known priors. Hence the higher redshift solutions identified by COSMOS2020 are thrown into significant doubt by the \bagpipes\ fitting result. If these galaxies are indeed at lower redshift, then they could cause over-predictions of temperature in samples based on the COSMOS2020 redshift estimates.

Those galaxies which show mismatched photometric redshift estimates also have lower \bagpipes\ mass predictions than had been estimated by COSMOS2020.
Several galaxies have mass estimates in the range $\mathrm{log(M}/\mathrm{M_{\odot}})=5-6$, which may be surprising as more typical galaxies at these redshifts have $\mathrm{log(M}/\mathrm{M_\odot})>8.$ This may suggest that \bagpipes\ has underestimated the redshifts, while COSMOS2020 has overestimated them. In these cases, an intermediate redshift is perhaps more likely than either of the derived values. This may imply that the simplistic star formation history and metallicity prescription used in the \bagpipes\ fits, and the photometric analysis and stellar population tracks used in the COSMOS2020 fits, are both inappropriate. However visual inspection shows that in the majority of cases, galaxies with a large photometric redshift discrepancy are in fact those with more limited data or missing photometric bands.

Deriving redshifts from photometric observations can be problematic since multiple templates can equally (or nearly-equally) fit the photometric observations. Where the posterior probability distribution of the photometric redshift fit is multimodal, with both high and low redshift solutions, the low redshift solution is generally preferred due to the strong evolution in source detectability and the steep luminosity function that disfavours bright galaxies in the distant Universe. Many photometric redshift fitting codes are designed to work on large surveys with rapid fitting algorithms that do not fully sample the distribution posterior. Others require presumptions regarding the distribution priors. Throughout we have used the COSMOS2020 redshift derived from the relatively-simple {\sc{LePhare}} algorithm. We discuss photometric redshifts within the COSMOS2020 catalogue further in appendix~\ref{sec:photo_append}.

Given the uncertainties in photometric redshift methodology, the cleanest samples will be derived from candidates in which multiple methods suggest a high redshift solution, while no plausible low redshift secondary solution exists. This will be a fairly rare occurrence.  By contrast, any disagreement between  two  redshifts derived from well-established and tested photometric fitting codes throws into question the reliability of any one of the galaxy's redshift estimates, and flags it as a possible source of contamination in high redshift bins.

To remove any potentially mis-classified galaxies, we exclude from the sample any sources where the COSMOS2020 and \bagpipes\ redshifts differ by more than 0.3, or where the \bagpipes-derived redshift has an uncertainty greater than 30\%.
In combination with the photometric confusion cut already described, removing these galaxies further reduces the sample  to 2113 galaxies.
We define this as our \textit{gold sample}, containing as little contamination as possible.

\begin{figure}
    \centering
    \includegraphics[width=\columnwidth]{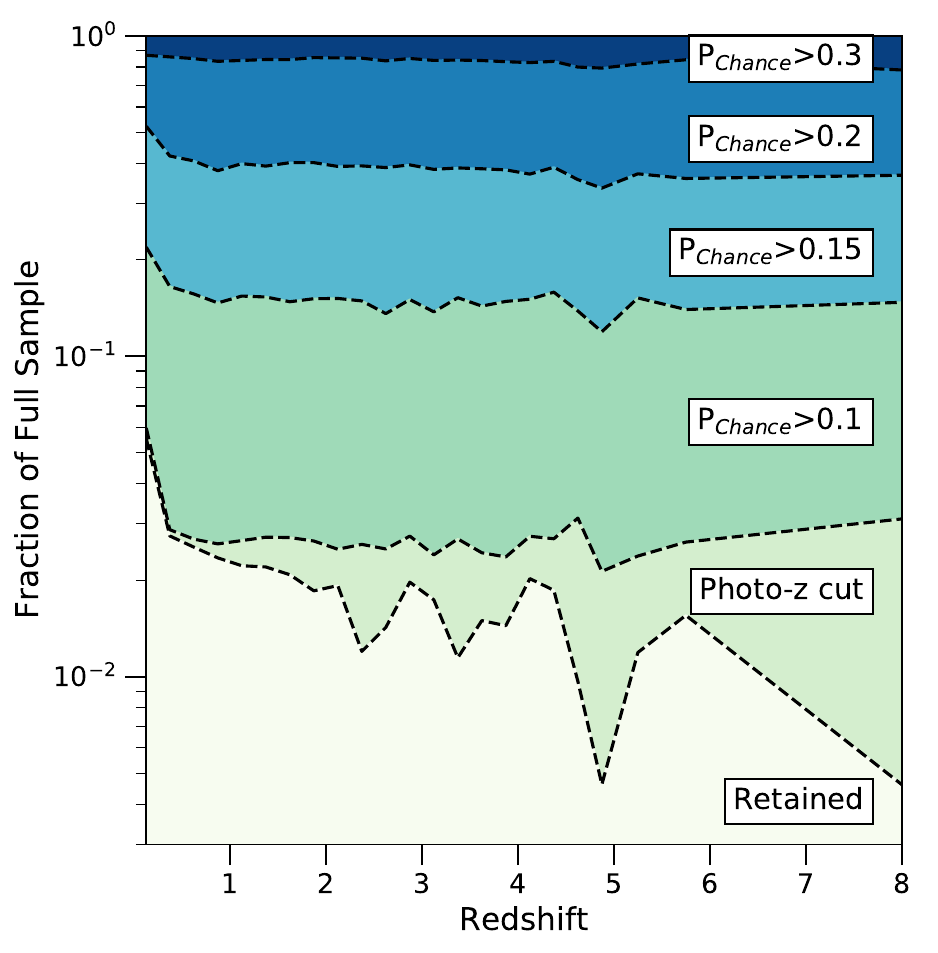}
    \caption{Fraction of galaxies remaining compared to the full sample for each redshift bin, plotted at the middle of the bin's width, after doing various cuts on the sample to remove contamination. The probability of chance alignment, $\mathrm{P_{Chance}}$, are shown for a number of cut values where everything above each dotted line has a $\mathrm{P_{Chance}}$ value greater than the one labelled. The photo-z robustness test is only done on galaxies which have $\mathrm{P_{Chance}}<0.1$ and any galaxy with a mismatching redshift between the \bagpipes\ and COSMOS2020 derived values lies above the ``phot-z cut" line. Any galaxy below is retained within the ``gold" sample.}
    \label{fig:frac_remain}
\end{figure}

In Fig.~\ref{fig:frac_remain} we show the fraction of galaxies remaining compared to the full sample after applying the probability of chance alignment and photometric redshift cuts.
We show the impact of cutting the sample on various values of $\mathrm{P_{Chance}}$. We then further illustrate the impact of the photo-z test on galaxies with $\mathrm{P_{Chance}}<0.1$.
The chance alignment cuts reduce the galaxy numbers almost independently of redshift at all $\mathrm{P_{Chance}}$ values. There is no clear bias towards rejecting a larger fraction at some redshifts compared to others. This suggests that the majority of contamination occurs from sources in the very local Universe ($z<0.25$), and almost homogeneously across the field. 

The photometric redshift cut, however, does have a redshift dependence with more galaxy candidates being lost at higher redshifts.
As mentioned, confirming high-z galaxies using photometry is difficult due to decreasing fluxes and low-z, dusty, red galaxies mimicking the properties of the high-z galaxies allowing for multiple redshift solutions. This latter can be observed at specific redshifts which have increased fractions of galaxies with large photometric redshift uncertainties due to confusion between spectral features when fitting (i.e. the Balmer break being mistaken for the Lyman break).

From this point forward, we use a stringent cut of $\mathrm{P_{Chance}}<0.1$.
Substituting $\mathrm{P_{Chance}}<0.2$ would retain almost half of the full sample, however 1-in-5 of these galaxies would likely be a contaminant, or roughly 10,000 galaxies could be contaminants. Since these contaminants would be dominated by local sources, they would likely be disproportionately bright in the IR, and their dust emission observed at shorter wavelengths than the high redshift targets, and so would disproportionately bias the measurements.

By selecting $\mathrm{P_{Chance}}<0.1$, we restrict our analysis to a small fraction of the original galaxies (2944, before photo-z cut), but would only expect roughly 300 of these to be contaminants. By further requiring that two indpendent techniques agree unambiguously on the photometric redshift of each target, we also remove remaining contaminants, which are disproportionately likely to be low redshift, dusty (and therefore IR-luminous) sources.

 The remaining sample of galaxies in any redshift-mass bin should not be biased by either of these foreground-removal cleaning techniques, and will allow derivation of a dust temperature relation less likely to be biased towards unphysically high temperature estimates than the full sample. 

\subsection{Gold Sample Analysis and Results} \label{sec:results}
The gold standard sample are binned and fit for temperature using the same methodology as the original sample, removing any redshift-mass subsample for which there is not enough IR detections from \simstack. 
As was the case with the full sample, we see two populations: those with low dust temperatures and others with higher estimated temperatures and/or larger uncertainties resulting from poor fitting. As the size of each sample has been dramatically reduced, more subsamples fall in the latter category.

\begin{figure}
    \includegraphics[width=\columnwidth]{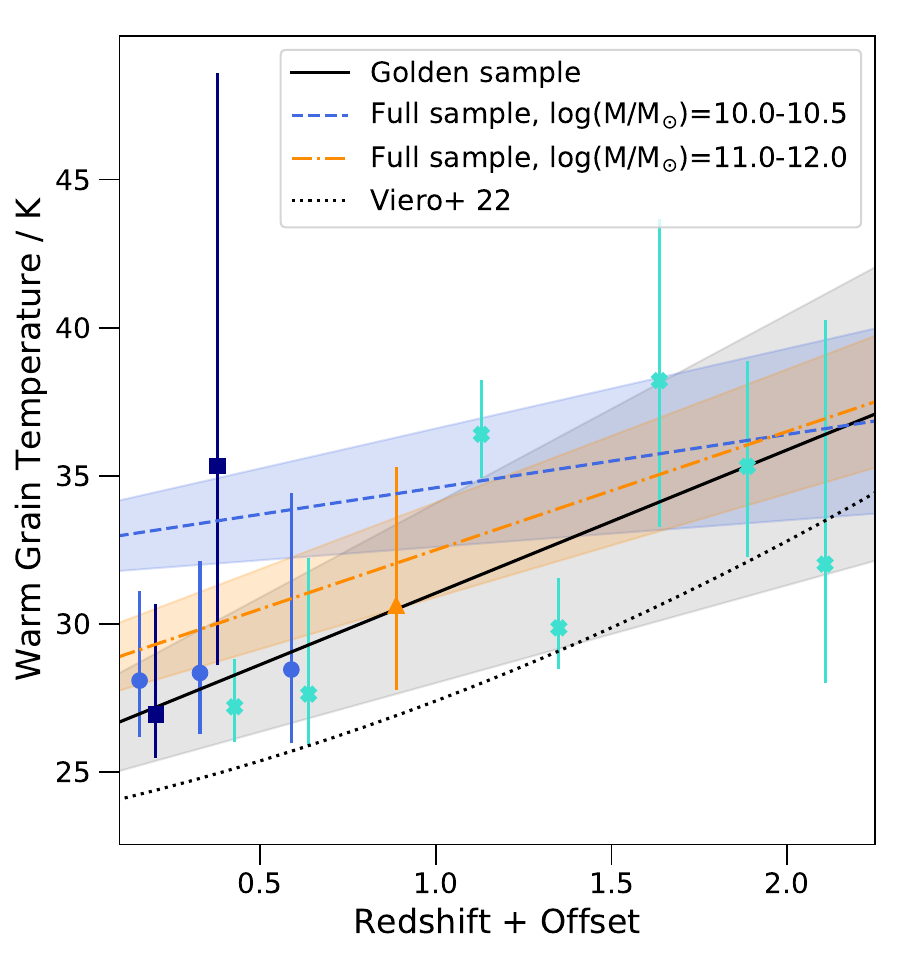}
    \caption{Derived dust temperatures for the golden sample after removing all samples with poor data and model fits. The symbol convention is the same as in Fig.~\ref{fig:temp_inc_bins}. The medium blue circle data points for the mass bin $\mathrm{log}(M/\mathrm{M_{\odot}})=10.0-10.5$ have been offset by $-0.05$ in redshift for clarity. Data across all bins has been fitted simultaneously using a linear least-squares to derive a $T_d-z$ relation of $T(z) = (4.8\pm1.5) \times z + (26.2\pm1.5)\ \mathrm{K}$, shown as a solid black line. The dashed-dotted orange and dashed blue lines are the best-fits for the mass bins $\mathrm{log(M}/\mathrm{M_{\odot}})=10.0-10.5$ and $11.0-12.0$ from the full sample, as given in Table~\ref{tab:best_params}. The shaded regions correspond to one sigma limits on each of the derived relations. The derived relation from \citet{2022MNRAS.516L..30V} is shown as a black dotted line for comparison.}
    \label{fig:temp_gold_samp_vet}
\end{figure}

We vet all best-fit models by the same criteria as before, and show the remaining temperature estimates in Fig.~\ref{fig:temp_gold_samp_vet}.
This sample is then fitted as before to derive,
\begin{equation}
 T_d(z) = (4.8\pm1.5) \times z + (26.2\pm1.5)\ \mathrm{K}.
\label{eqn:temp2}
\end{equation} 
This relation is slightly steeper than the ones found for the original sample, but is in agreement with those found for the higher two mass bins. This may result in part from the limited data which is biased towards low mass galaxies at low redshift, and high mass galaxies at high redshift, introducing an artificial steeping of the relation. An interesting result is that the gold sample selection has disproportionately removed measurements with high dust temperatures, resulting in a lower overall normalisation of the relation derived. This suggests that the majority of the highest derived temperatures resulted from galaxies with confused, or otherwise poor, photometric data and that the intrinsic dust temperature of the population in the redshift range reliably accessible to the COSMOS2020 survey is low.

\begin{figure}
    \includegraphics[width=\columnwidth]{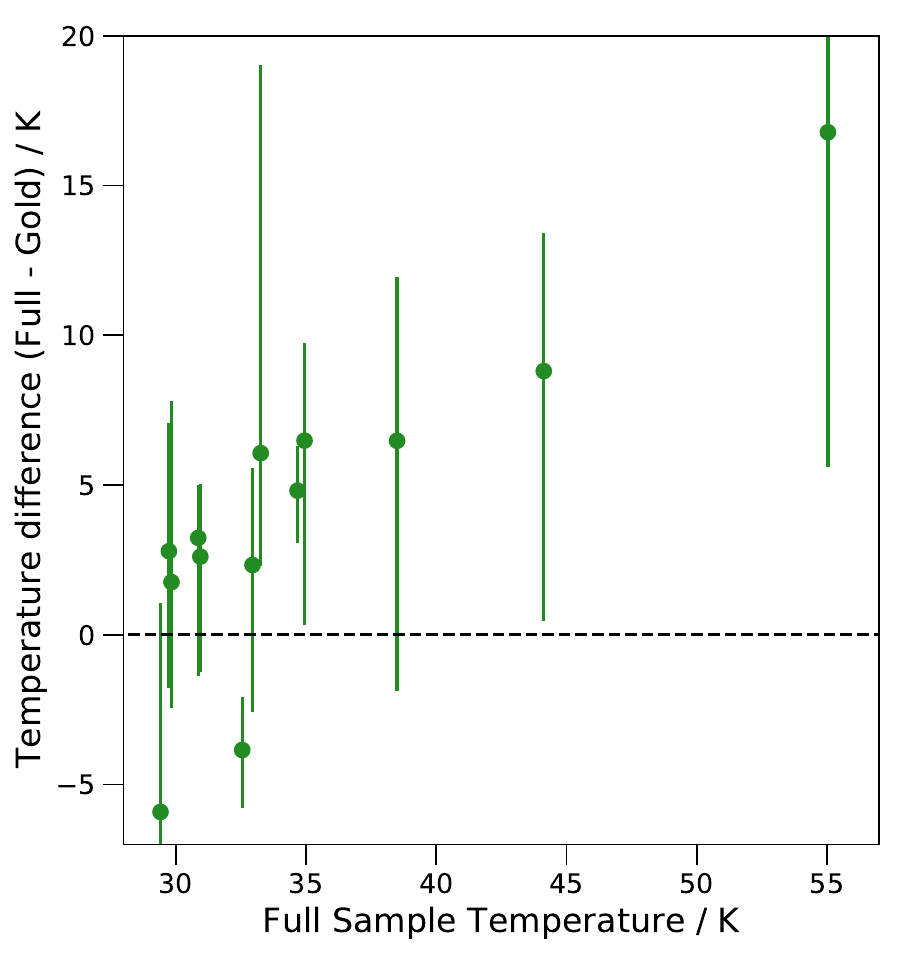}
    \caption{Comparison of the derived temperature for the subsamples from both the full and golden samples. A positive trend can be seen, with the higher derived temperatures from the  full sample having a larger difference to that derived for the same subsample in the gold sample. The mean difference between the samples is $4.0^{+5.0}_{-1.9}$\,K. Note that the most left data point has a lower limit of $-19.3$\,K and the most right data point has an upper limit of $78.8$\,K.}
    \label{fig:samp_temp_comp}
\end{figure}

In Fig.~\ref{fig:samp_temp_comp} we compare the derived temperature from the full and golden samples on a case by case basis for redshift-mass bins which are populated in both analyses.
The majority of subsamples have higher derived temperatures in the (contaminated) full sample, and the difference becomes larger the higher the derived full sample temperature. This suggests that, as expected, contamination is causing the apparent mean temperature to increase in the stacked data.
The average difference between the two samples is $\Delta$T$=4.0^{+5.0}_{-1.9}$\,K.

The difference between full and gold samples could, theoretically arise either from the simple contamination from line-of-sight coincidence, or from catastrophic failures in estimated redshift adding contaminants to the target sample. 

At low redshifts ($z<1$), the golden selection removed very few sources due to photometric redshift constraints, and more due to confusion. The existence of a temperature difference in this redshift regime thus implies that the screen of coincident low redshift foreground sources typically acts to raise the derived dust temperature estimate.

At higher redshifts the origin of the apparent offset is less clear, as both line-of-sight coincidence and sample contamination due to poor photometric redshift estimates  significantly affect the sample size. Since dust emission from low redshift interlopers will peak at short wavelengths, implying a higher apparent temperature for contaminated samples than clean samples, it is likely that both biases work in the same direction and the high dust temperatures measured in the full sample are unreliable. The implied increase in temperature offset with temperature in Fig.~\ref{fig:samp_temp_comp} is thus unsurprising.

Given the direction of the biases, we take the dust evolution from the full sample to act as an upper limits for the average dust temperature evolution with redshift. We note that we derive $T_d-z$ relations with similar slopes in both the full and gold samples and so that the effect of contamination (line of sight or interloper) on the full sample manifests as an offset to higher temperatures, rather than a severe modification of the relationship slope.

\section{Discussion} \label{sec:discussion}

\subsection{Constraining the Dust Temperature Evolution} \label{sec:temp_discussion}

\subsubsection{Observational evidence for dust temperature in the literature}
In the distant Universe, studies of individual galaxies are relatively limited in number and have estimated a wide scatter of dust temperatures.
\citet{2022MNRAS.515.1751W} find that the Lyman-break galaxy COS-3018555981, at $z=6.854$ (where $T_{\mathrm{CMB}}=21.40$\,K), has a dust temperature of 26.24\,K with a upper limit of <\,48\,K (95\% confidence). This is barely above the CMB temperature at this redshift, and would imply a very limited increase in temperature with redshift, as has been suggested previously \citep[e.g.][]{2019MNRAS.489.1397L, 2022ApJ...930..142D}.
\citet{2022MNRAS.515.1751W} also find two other Lyman-break galaxies at a similar redshift to have slightly higher temperatures but with very large uncertainties ($41.8^{+35.4}_{-15.1}$\,K and $53^{+36.8}_{-17.5}$\,K) arising due to estimates being calculated using only two flux measurements.
The analysis by \citet{2020MNRAS.498.4192F} find an average dust temperature of $38\pm8$\,K at $z\sim5.5$ for four main-sequence galaxies, which is also below what is expected by extrapolation of trends at lower redshifts.
By contrast, estimates of very high dust temperatures have been suggested by \citet{2018MNRAS.477..552B} for the dusty galaxy A2744\_YD4 at $z=8.38$ ($T_d=91$\,K) and by \citet{2020MNRAS.493.4294B} for the Lyman-break galaxy MACS0416\_Y1 at $z=8.31$ ($T_d>80$\,K, 90\% confidence limit).
This would instead require a steep temperature dependence on redshift \citep[e.g.][]{2022MNRAS.516L..30V}.

Apart from \citet{2018MNRAS.477..552B}, all these estimates were calculated by fitting to SEDs with limited data ($\le4$ data points).
For the dusty galaxy A2744\_YD4, \citet{2017ApJ...837L..21L}, who did the original analysis, only found a dust temperature in the range $37-63$\,K when fitting its SED using MAGPHYS \citep{2008MNRAS.388.1595D, 2015ApJ...806..110D}.
However, this analysis was limited by the use of only one filter in the far-IR (a synthetic filter of the ALMA band~7).
\citet{2018MNRAS.477..552B} reanalysed the data, comparing to hydrodynamic and dust radiative transfer simulations.
Having limited data for this galaxy is the major issue for determining the true dust temperature accurately.
To improve the parameter estimates, ALMA band~5 and~6 observations were taken.
This resulted in non-detections of the dust continuum, constraining the dust temperature to $>$\,55\,K ($>$\,43\,K) using $2\sigma$ ($3\sigma$) upper limits \citep{2019MNRAS.487L..81L}.
It should also be noted that directly fitting SED observations yields a luminosity-weighted dust temperature, which can skew the temperatures higher due to hot dust in star-forming regions \citep{2019MNRAS.489.1397L}.
\citet{2018MNRAS.477..552B} do find a much lower dust temperature of $T_d=50\pm19$\,K when calculating a mass-weighted mean in their simulations, which is in agreement with the \citet{2017ApJ...837L..21L} value.

Larger samples have been considered by a handful of surveys, mostly based on data obtained using the full sensitivity of ALMA.
These include the ALMA Large Program to INvestigate [CII] at Early times \citep[ALPINE,][]{2020A&A...643A...2B} and the Reionization Era Bright Emission Line Survey \citep[REBELS,][]{2022ApJ...931..160B} programs, both of which incorporate subsets of the COSMOS field.
These are ALMA Large Programs, designed to target bright [CII]$_{158\mu\mathrm{m}}$ lines and dust-continuum emission from UV-selected galaxies at $4<z<6$ and $z>6.5$ for the ALPINE and REBELS surveys, respectively.

\citet{2021MNRAS.503.4878S, 2022MNRAS.513.3122S} proposed a novel method to derive dust temperatures in the REBELS survey, using a combination of the 158\,$\mu$m continuum and the super-imposed [CII] line emission, since they only have single far-IR data points which leaves SEDs unconstrained.
The same method is used in \citet{2022MNRAS.517.5930S} on the ALPINE galaxies.
\citet{2022MNRAS.512...58F} constrains the dust temperature of the REBELS galaxies by using the UV spectral slope, the observed UV continuum flux at 1550\,\AA, and the observed far-IR continuum flux for each galaxy in their sample.
The derived dust temperatures from these analyses, along with temperatures derived using modified blackbody fits to SED stacks of ALPINE-like galaxies at both $4<z<5$ and $5<z<6$ \citep{2020A&A...643A...2B}, are plotted in Fig.~\ref{fig:dust_temp_comp}, shown as the average temperature for the galaxies in the program.
Both independent methods for fitting the ALPINE and REBELS galaxies have derived consistent average dust temperatures for these samples.

In the local-to-intermediate redshift Universe ($z<4$), \citet{2023arXiv230506388M} report the dust temperature in 57 dusty star-forming galaxies.
These are 870~$\mu$m selected galaxies from the GOODS-S field.
They use an optically thin, modified blackbody to constrain SEDs comprised of Herschel/PACS, Herschel/SPIRE, SCUBA-2, and ALMA data, giving up to 10 flux measurements that are used for fitting.

\subsubsection{Previously derived trends}
Previous analysis of trends in the dust temperature evolution with redshift have suggested a linear dependence of temperature on redshift of $T_d=(32.9\pm2.4)+(4.60\pm0.35)\times(z-2)$ \citep[][]{2018A&A...609A..30S} and $T_d=(34.6\pm0.3)+(3.94\pm0.26)\times(z-2)$ \citep[][]{2020ApJ...902..112B}.
The sample in \citet{2018A&A...609A..30S} uses the GOODS-South, GOODS-North, UDS and COSMOS fields out to $z\sim4$, while the sample in \citet{2020ApJ...902..112B} uses star-forming galaxies at $z=1.5-10$ in the Hubble Ultra-Deep Field.

However, \citet{2022MNRAS.516L..30V} found the dust temperature to increase as a second-order polynomial with redshift.
This is consistent with the outliers of \citet{2018MNRAS.477..552B} and \citet{2020MNRAS.493.4294B}, and predicts $T_d=105$\,K at $z=8.38$.
\citet{2022MNRAS.516L..30V} use the same COSMOS2020 galaxy photometric redshift-selected sample as in our full analysis, without making any substantial cuts for completeness, contamination or wavelength coverage. They also use \simstack\ to generate the far-IR measurements in the same bands.

\subsubsection{Reconciling our results with previous work}
\begin{figure}
    \includegraphics[width=\columnwidth]{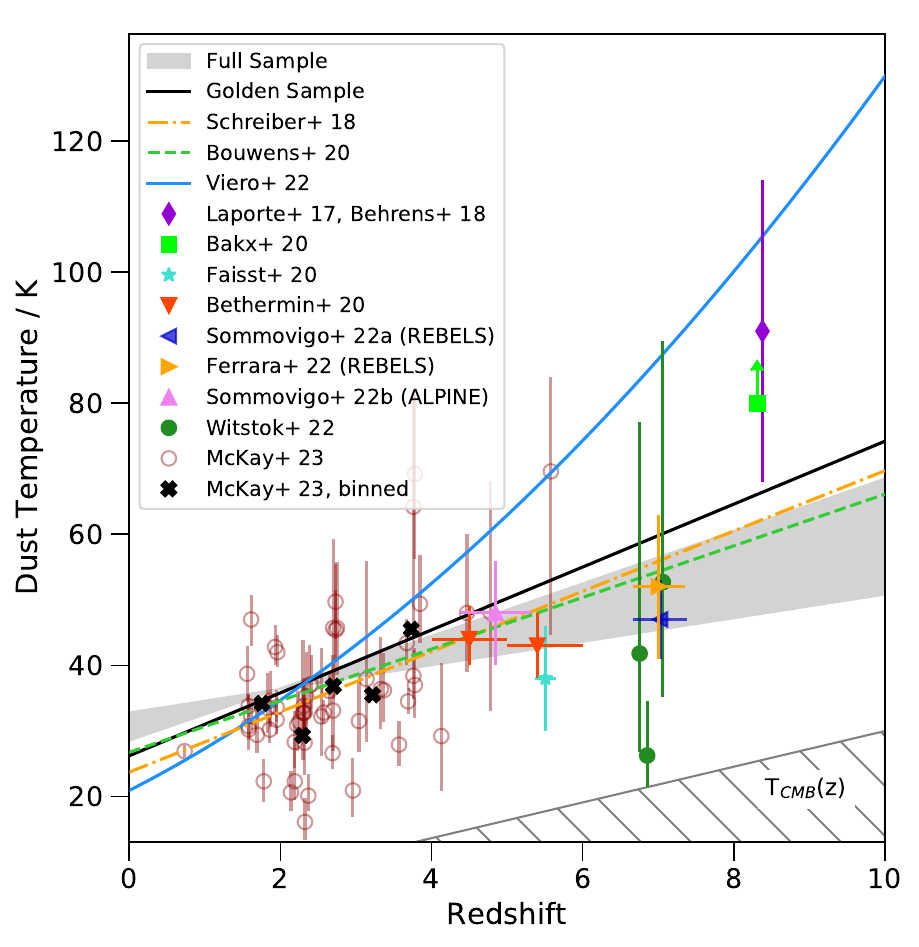}
    \caption{The derived evolution of dust temperature with redshift in the context of previous work. The shaded area represents the upper and lower bounds for the dust temperature evolution from our full sample analysis. The area is bounded by our shallowest slope from fitting the mass bin $\mathrm{log(M}/\mathrm{M_{\odot}})=10.0-10.5$ and the steepest from $\mathrm{log(M}/\mathrm{M_{\odot}})=11.0-12.0$, extrapolated to $z=10$.
    The solid black line is the best fit from our golden sample for reference, also extrapolated.
    Lines indicate relations derived by other authors; \citet{2018A&A...609A..30S} as the dot-dashed orange line (extrapolated from $z=4$), \citet{2020ApJ...902..112B} as the dashed green line, and \citet{2022MNRAS.516L..30V} as the curved solid blue line. Points indicate individual temperature estimates at fixed redshift; the purple diamond for the dusty galaxy A2744\_YD4 \citep{2017ApJ...837L..21L, 2018MNRAS.477..552B}, green square for the lower limit of Lyman break galaxy MACS0416\_Y1 \citep{2020MNRAS.493.4294B}, light blue star for the average of four main-sequence star forming galaxies \citep{2020MNRAS.498.4192F}, red down-pointing triangle for the two stacked groups of ALPINE-like galaxies in the COSMOS sample \citep{2020A&A...643A...2B}, dark blue left-facing and orange right-facing triangle for the REBELS galaxies \citep[][respectively]{2022MNRAS.513.3122S, 2022MNRAS.512...58F}, the pink up-facing triangle for the ALPINE galaxies \citep{2022MNRAS.517.5930S}, green circles for three Lyman-break galaxies \citep{2022MNRAS.515.1751W}, and red open circles are for the 57 dusty star-forming galaxies \citep{2023arXiv230506388M} with the thick black crosses showing the average binned with spacing $\Delta z=0.5$ between $1.5<z<4$ (errorbars not shown for clarity). The CMB temperature evolution with redshift is shown in the lower right and is expected to act as a lower limit for the temperature of sources in equilibrium with their surroundings.}
    \label{fig:dust_temp_comp}
\end{figure}

In Fig.~\ref{fig:dust_temp_comp}, we compare our derived dust temperature evolution with redshift to these previous analyses.
Also plotted is the evolution of the CMB temperature.
All the derived $T_d-z$ relations are consistent with the observed low, $T_d<$\,45\,K dust temperatures in the local Universe ($z<3$).

Our results have demonstrated the large amount of uncertainty produced when using photometric redshift derived samples to explore high-redshift ($z>4$) dust temperatures, and suggest that this has been underestimated in some past work. SED fitting requires robust knowledge of contamination in the sample, with sufficient data points spanning key wavelength ranges in the IR to constrain the peak and slope of the dust emission curves. From fitting both the full and golden samples in Figs.~\ref{fig:temp_inc_bins_vet_cut} and \ref{fig:temp_gold_samp_vet}, we have found that a linear increase in temperature with redshift can model the derived dust temperatures up to $z\sim5$ for the full sample (albeit with significant concerns regarding contamination) and up to $z\sim2$ for the golden sample.
The golden sample only extending to $z\sim2$ highlights the difficulty of getting clean, robust photometric samples of high redshift galaxies to derive average dust temperatures in the distant Universe, and this does call into question the reliability of the relations derived using larger photometrically-derived galaxy samples.

The golden sample was consistent with a similar redshift dependence but slightly lower temperatures (by $\sim4$\,K on average) at $0<z<2$ than the full sample, as seen in Fig.~\ref{fig:samp_temp_comp}.
If this continued to be the case in higher-$z$ but more contaminated samples, the estimated temperature would fall above the true temperature. It was highlighted in Fig.~\ref{fig:model_grid_5} and Section \ref{sec:contamination} that contamination in the UV and optical data was found in the full samples and as such, flux contamination was expected to be present in the infrared.
Contamination of low-z galaxies within the sample would result in higher predicted dust temperatures than the true values, as there is a degeneracy between the effect of redshift and temperature on the peak of the dust emission grey-body. An observed peak in the blue bands of the FIR can be interpreted either as a high temperature at high redshift, or as a low temperature at lower redshift. Thus our steepest fit to the full sample (found in $\mathrm{log(M}/\mathrm{M_{\odot}})=11.0-12.0$) in Fig.~\ref{fig:temp_inc_bins_vet_cut} acts as an upper limit for the average dust grain temperature at each redshift.

Our derived dust temperature evolution out to $z\sim5$ nonetheless agrees with those found using samples taken from multiple fields on the sky as Fig.~\ref{fig:dust_temp_comp} demonstrates. In particular it is consistent with the deeper observed constraints from the REBELS and ALPINE surveys at higher redshift. 
This suggests that the extrapolation of our best-fits up to at least $z\sim8$ are suitable for the average dust temperature evolution in the average star-forming galaxy in the sample.
Our predictions also agree with the dusty star-forming galaxies reported in \citet{2023arXiv230506388M} at lower redshifts ($1.5<z<4$) based on observations from a different field.
Their averages, shown as black crosses in Fig~\ref{fig:dust_temp_comp}, lie in agreement with the fits to both the full sample and that extrapolated from the golden sample derived here.
There is, however, a large scatter seen within the \citet{2023arXiv230506388M} sample, with most individual galaxies having derived temperatures between $15<T_d<50$\,K.
This highlights the diversity of galaxies which can be found within any sample, the complex nature of dust emission and the impact of variations in star formation history or metal enrichment.
Any $T_d-z$ relation based on the mean properties of galaxies will not be able to predict the dust temperature of any individual galaxy, however it should describe the average dust temperature of any sample at a given redshift.
This is what we see from our derived dust temperature relation, with agreement to the average of many samples.

The observed trend for both full and golden samples differs, however, from that derived by \citet{2022MNRAS.516L..30V} which, crucially, is based on the same data set considered here but without a careful assessment of the sample quality. Fig.~\ref{fig:dust_temp_comp} shows both the golden sample (black) and \citet{2022MNRAS.516L..30V} (blue) relations. These not only show a large disagreement beyond $z\sim6$, but are also inconsistent with one another in the redshift range $0<z_\mathrm{phot}<1.5$ where the golden sample provides constraints, highlighted in Fig.~\ref{fig:temp_gold_samp_vet}.  Thus the golden sample is not in agreement with the \citet{2022MNRAS.516L..30V} trend either at high redshifts or in the range supported by data. We cannot rule out an exponential trend at redshifts above those we have data for, however neither do we find any evidence to support it, and a linear trend provides a good fit to the data. If such an exponential trend exists, it would have to be differently normalised in the $0<z_\mathrm{phot}<2$  range than the \citet{2022MNRAS.516L..30V} relation. 

Setting aside the question of contamination, our methodology differs from that of \citet{2022MNRAS.516L..30V} as we use a finer redshift grid spacing and fit UV-optical data in addition to the IR. As discussed above, we expect the majority of the difference in estimated $T_d-z$ relation to result from the initial sample selection rather than fitting methodology. The very steep dust temperature evolution derived by \citet{2022MNRAS.516L..30V}  is driven by high estimated dust temperatures in the distant Universe. We do not see these, largely because we were unable to identify sufficiently robust subsamples in the COSMOS2020 dataset to enable fitting in this redshift regime. 
\citet{2022MNRAS.516L..30V} provides some discussion and calculations on the impact low-$z$ interlopers may have on their stacked SED shape, suggesting that the overall bias is small but introduces large scatter at $z>5$.
To account for this, they estimate excess uncertainties on the dust temperature estimates.
However, it is at these redshifts where the deviation between the \citet{2022MNRAS.516L..30V} predicted relation and the relation derived from our analysis becomes significant.
From our analysis, we suggest that the data in the COSMOS-derived photometric redshift sample above $z\sim5$ is significantly affected by both interlopers and line-of-sight coincidence and thus the data are too impure to fit.

We do, however, find good agreement with other previously derived trends, and with some individual galaxies. There remain low- and high-temperature outliers in the observational sample. The shallowest relation derived in our work (predicting $T_d=45$\,K at $z\sim6.8$) may be consistent with the \citet{2022MNRAS.515.1751W} $2\sigma$ upper limit. However, the measured value lies well below all extrapolations, relations, and estimates reported here or elsewhere. The estimated temperatures of A2744\_YD4 \citep{2018MNRAS.477..552B} and MACS0416\_Y1 \citep{2020MNRAS.493.4294B} are both above even our steepest relation (which yields 62\,K at $z=8.38$).
We do not exclude the possibility of these individual galaxies having extremely high or low dust temperatures, but this is not seen within our analysis and thus may is not typical for galaxies at these redshifts.

This scatter in the observed population highlights that extant dust temperature modelling may be oversimplistic and fail to account for galaxies with non-standard properties (i.e. decreased dust abundance in some high-$z$ galaxies or intense star formation).

Ultimately, only as more data, which is free from contamination, is collected will we find out the true dust temperatures for galaxies in the high-$z$ regime and thus, the average evolution.

\subsection{UV-IR Spatial Disconnect} \label{sec:disconnect}
The \bagpipes\ fitting program underlying the above analysis is built on a key assumption: that the stellar population identified in the optical is responsible for the far-IR dust emission, i.e. that the energy-balance formalism can be applied. Since the number and precision of UV-optical datapoints tightly constrains the fitting procedure (see Section~\ref{sec:bagpipes}), the IR emission does not drive the energy-balance estimate. 

The best-fitting model outputs from \bagpipes\ can be classified into three empirical categories; stacks with good fits to all or most data points with low derived temperatures and errors (i.e. top panel in Fig.~\ref{fig:model_grid_5}), stacks with missing 160~$\mu$m fluxes (i.e. second panel), or stacks with IR data but for which \bagpipes\ stellar population synthesis fitting under-predicts the amount of IR flux required (i.e. bottom three panels).
Whilst the first two categories occurred at all redshifts, the final category only occurred when fitting samples at $z>4.5$.
15 out of 20 samples above this redshift had poor fits due to under-predicting the IR flux.
Whilst a warmer dust grey-body component could fit the observational data, this is not consistent with the stellar population and its constraints on the dust properties.
This occurs due to two issues: either there are too many interlopers within the sample or there is a disconnect between the stellar populations giving rise to UV-optical and IR emission in high redshift galaxies.

Studies involving high spatial resolution observations at submilimetre and millimetre wavelengths have found there to be spatial offsets between the peak emission in the rest-frame UV and far-IR emission, at both high-$z$ \citep[$z>6.5$, e.g.][]{2015MNRAS.452...54M, 2017A&A...605A..42C, 2022MNRAS.515.3126I} and intermediate-$z$ \citep[$z=3-5$, e.g.][]{2016ApJ...833..103H, 2019ApJ...882..107R, 2021MNRAS.503.2622C}.
Even in the local Universe, studies have found inhomogeneities in the distributions of the UV-optical and IR emission. However, a detailed investigation of spatially resolved emission in the interacting galaxy Arp 244 has shown that such inhomogeneities  do not affect the globally-averaged performance of SED fitting using an energy-balance formalism \citep[][]{2022A&A...665A.137S}.

This is a little surprising, as it implies that although the dust is not co-located with the stars it is nonetheless heated by them, with no additional dust-obscured or otherwise hidden heat sources. 
Instead, if the locations where UV-optical and infrared emission originate in the galaxy are separated, then the two emissions could be produced from two populations, i.e. the dust is being heated by a different population than the one seen in the UV-optical. In this case, global SED fitting through the energy-balance formalism should fail.

There are relatively few local galaxies sufficiently resolved for such studies, and similar analyses are impossible on high redshift galaxies themselves. If the interaction geometry or dust extinction distribution is very different in distant galaxies, then highly extincted regions may not be represented in the UV-optical data. In this case the energy-balance relation may break down and the observed UV-optical emission might not produce enough energy to balance the IR emission.

Although the analysis in our work provides some evidence for such a failure of energy-balance assumptions, we caution that low redshift contamination of the sample remains an alternate explanation.
Lower redshift galaxies of a fixed mass will appear brighter than their high redshift counterparts, meaning they can bias inferred luminosities.
The observed angular resolution in the infrared can also have the same effect.
Interlopers and contamination could manifest in three primary ways: (i) both UV-optical and IR photometry may be contaminated, (ii) the UV-optical may be contaminated with the IR measurement remaining robust, or (iii) the IR may be contaminated with the UV-optical uncontaminated.

Option (i) may be true for some subsamples as 11 of the 20 purely photometric redshift based galaxy subsamples above $z>4.5$ have clear detections in the UV-optical data that would lie below the presumed Lyman limit. This strongly implies the presence of interlopers.
This excess UV-optical emission is not expected in high-$z$ stellar population models which incorporate absorption in the intergalactic medium. Thus a lower redshift source of ultraviolet photons must be present. Since we observe its UV flux below the Lyman limit, we must also infer a reprocessed IR emission component, but at an unknown redshift and wavelength. Since this does not appear within the models, which assume the photometric redshifts are reliable, an energy-balance formalism cannot work.

In Fig.~\ref{fig:samp_temp_comp}, we showed that restricting contamination in the local Universe by requiring a high fidelity in the data reduces the dust temperature. This suggests that the net effect of contaminating flux is to bias the results to higher temperatures. 
Thus, if  observations beyond $z\sim4.5$ that show rest-far-UV emission are contaminated, then no information can be derived from this sample and only extrapolations from cleaner, lower-$z$ samples can be used. 
However, interlopers may not explain all failures of the  UV-optical stellar population to explain the IR emission, since UV-flux contamination is not always detected.

In scenario (ii), laid out above, only the UV-optical emission is contaminated, with the contaminants in starlight contributing relatively little extra stellar radiation from hot stars for dust reprocessing. This is possible as not all stellar populations produce strong IR emission. However, blending and source confusion are more of an issue due to large PSFs in the infrared, and the contaminants will typically be at lower redshifts (and thus easier to detect in the IR) than the target galaxies, regardless of their stellar population. This may lead to lensing of the IR flux from the background source \citep[e.g. ][]{2002MNRAS.329..445P} or IR flux contamination. It is thus relatively unlikely that the IR can escape contamination, if UV-optical flux is significantly contaminated.
 
Case (iii) suggests that there could still be contamination in the IR observations in the 9 high photometric redshift subsamples which have no clear UV-optical contamination.
In this scenario, the fits to the UV-optical data produce a physically motivated dust emission which falls short of the observed IR emission, since the excess IR emission can be attributed to contamination by sources which are heavily extincted in the UV-optical, or IR-luminous emission mechanism which are not accounted for in our model, such as AGN.
However, this UV-optical faint component could instead be a highly-obscured stellar component 
associated with the target but 
spatially offset to the peak UV-optical emission.
Thus, case (iii) suggests that the energy-balance derived FIR SED would sometimes underpredict the observations and imply the presence of contamination or a non-stellar dust heating component.
Only a cleaner sample would be able to disentangle the degeneracy between these two possible explanations.

A final interpretation (scenario iv) could be that the infrared emission models adopted are inappropriate. However this is not supported by the analysis, as the emission component in the far-IR can  be independently fit both by \simstack\ and \bagpipes\ when the UV-optical data is excluded.
The issue here is instead the relation between the UV-optical data and the infrared. 

The observed trends in the data suggest that both UV-optical and IR components may be contaminated in many cases in the full sample. In the gold sample we have attempted to minimise potential for scenarios (i) and (ii), leaving scenario (iii) - a UV-optically faint component as the preferred interpretation. However this sample does not extend to the redshifts where a  UV-optical/IR disconnect is seen in the SED fitting. As a result it cannot be used to distinguish between the three scenarios, each of which is likely to be contributing in some part.
Only with a spectroscopically confirmed, high spatial resolution, contamination-free sample will the origin of the UV-optical/IR disconnect be fully resolved.

\subsection{Limitations in Dust Modelling} \label{sec:limits}
In Section~\ref{sec:simstack_temps}, we highlighted that use of different IR models causes variations in the derived dust temperature.
An average scatter of $\sim20$\,K was found, which is concerning given that the temperatures in the sample are between $30-50$\,K.
Even \citet[][]{2022MNRAS.516L..30V}, which did the same analysis but with a coarser redshift grid spacing, predict lower temperatures by $\sim10$\,K at $z=0$ and higher temperatures of $\sim20$\,K by $z\sim5$, showing this discrepancy.
\citet{2022MNRAS.514.5706J} found that the choice of stellar population synthesis model also affects derived dust temperature, with single star SPS models typically requiring higher temperatures compared to ones which include binaries. Thus, the use of BPASS models could be driving the dust temperatures slightly lower in this analysis than in some other work.
The choice of model used and the assumptions which go with it need to be considered to determine whether it is appropriate before any fitting is undertaken.

The \simstack\ software uses a very simple model assuming that the dust can always be modelled using a single blackbody curve and no other components are required. This is a common approach \citep[e.g.][]{2010MNRAS.409...75H, 2012MNRAS.425.3094C, 2022MNRAS.515.1751W}. Indeed, if unconstrained by the need for energy-balance with the stellar population, \bagpipes\ can also fit arbitrarily warm dust temperatures which may better reproduce the dust emission in some cases.
Whilst this approach allows a constraint to be determined from any IR SED, the model will miss the complex interplay between the nature of the dust and the stars occurring within galaxies.
Our analysis, on the other hand, has taken the opposite approach and used a model with many components, including a stellar emission component. This allows far more data to be considered in the interpretation (i.e. UV-NIR photometry), at the expense of having more parameters to constrain.
This approach will produce more physical constraints if a connection between the dust properties and stellar population is presumed \citep[][]{2021ApJ...917....3D}. 
It is limited, however, by the need of a comprehensive wavelength coverage, degeneracies between parameters, and assumptions about all UV, optical and IR components including how they are connected. It thus cannot be applied to large samples in which infrared data only is available. As we have argued, this should not introduce a bias in the temperature measurements, except in the case of optically thick dust, whose heating source is not represented amongst the observed stellar flux.

For this study, we assumed that the same range of dust model parameters can be applied regardless of redshift, which may not be appropriate.
Dust properties might be expected to change over cosmic time due to the different timescales of different dust formation and destruction mechanisms \citep[e.g. in supernovae or cool star winds,][]{2007MNRAS.378..973B,2011ApJ...727...63D,2019A&A...624L..13L}, altering the IR emission curve. However, studies  have found certain key inferred properties of dust, in particular the dust emissivity index, not to evolve \citep[e.g.][]{2023MNRAS.523.3119W}. By fixing the emissivity curves, while allowing the strengths and temperatures of different components to vary, 
the empirical model applied in this study should be sufficiently flexible to be suitable at the redshifts fitted.

As has been pointed out, a crucial property which goes into the IR model is the opacity regime of the dust.
In our analysis, by coupling the optical and far-infrared emission through an energy-balance formalism, we have implicitly assumed an optically thin dust opacity. Deriving the opacity individually for each galaxy is not possible especially when stacking these faint sources.
However, there are known galaxies where the dust opacity is optically thick and thus our assumptions would not be valid.
Studies investigating how dust opacity affects the derived dust temperature find that an optically-thin assumption may underestimate the dust temperature \citep[e.g.][]{2019ApJ...887..144J, 2022A&A...665A...3J, 2020A&A...634L..14C, 2020ApJ...895...81R, 2023MNRAS.523.3119W} if a hot irradiating source is embedded in dense gas and dust. Such sources would often be AGN, but could include extremely young starbursts. These would appear in our analysis as catastrophic failures of the fitting algorithm, which would be rejected from further analysis. Hence, our derived $T_d-z$ relation might under-predict the temperature evolution if these sources are overrepresented in the distant Universe.

In reality, it is likely that the sample will contain a mixture of optically thick and thin dust in galaxies, with the latter dominating.
Thus, whilst our relation will strictly be valid for optically-thin galaxies only,  more observations would be required to investigate the ratio between optically thick and thin galaxies and their dust temperature evolution. This is beyond the capacity of any extant telescope or survey.

\section{Conclusions} \label{sec:conclusions}

In this work we have explored the dust temperature constraints obtained from SED fitting of a large galaxy sample, derived from public survey data, spanning a redshift range $0<z<10$. In this analysis, we adopt an energy-balance formalism in which the dust properties are affected by the stellar irradiation, and fit the stellar component with spectral synthesis models incorporating the effects of binary evolution. We have explored both the physical inferences to be drawn from this population, and methodological limitations in drawing such conclusions.

We summarise our main results as follows:

\begin{enumerate}
    \item Galaxy stacks based simply on photometric redshifts appear to show bimodal dust temperatures. The low temperature population is more robustly constrained than the higher temperature population, which typically has fewer infrared detections around the expected peak of dust emission and shows evidence for flux contamination by interlopers. 

    \item For galaxies with stellar masses  $\mathrm{log(M/M}_\odot)>9.5$, the evolution in the dust temperature can be traced in photometric redshift samples from $z=0$ to $z\sim5$.
    These temperatures are typically between $30<T_d<45$~K and show a smooth increase to higher temperatures with increasing redshift.

    \item We find that insufficient wavelength coverage and sensitivity, and the frequent presence of interloping foreground galaxies in photometric redshift-derived samples, prevent robust temperature determinations at $z>4.5$ in all mass ranges. Where galaxy subsamples are well detected at high redshift, the energy-balance formalism frequently underpredicts the far-infrared flux, implying the presence of either contamination, a highly-extincted stellar population, or a non-stellar, dust-heating component (i.e. AGN).

    \item The results above are supported by a ``gold standard" sample selection designed to minimise redshift uncertainty and confusion from foreground interlopers. This probes $0<z<2$ and suggests that all samples at redshifts above this range should be treated with caution in the absence of 100 per cent spectroscopic redshift confirmation.

    \item Our dust temperature relation, $T_d(z)=(4.8\pm1.5) \times z + (26.2\pm1.5)\ \mathrm{K}$, agrees well with those determined by some previous authors, for example \citet{2018A&A...609A..30S}, \citet{2020ApJ...902..112B} and \citet{2022MNRAS.513.3122S,2022MNRAS.517.5930S}. By contrast, our results do not support the dust temperature-redshift evolution relation determined by \citet{2022MNRAS.516L..30V}, which would imply much higher temperatures in the high redshift regime.
    
\end{enumerate}

A robust determination of dust temperature from SED fitting requires a sample with both a reliable redshift and several high signal-to-noise photometric detections in a wavelength range that constrains the peak of the dust emission curve, which will lie between about 30 and 300 microns in the rest frame. Current large galaxy surveys with data from instruments such as Herschel have insufficient depth and spatial resolution to obtain such photometric information for typical distant star forming galaxies.
The most robust data currently comes from the Atacama Large Millimeter/submillimeter Array (ALMA), as was used in the ALPINE and REBELS programs.
ALMA's observing wavelengths are at $>300\,\mu$m, meaning that it can be used for high-$z$ (i.e. $z>3$) dust temperature measurements.
However, most of the observing bands 
do not constrain the peak of the blackbody emission at typical dust temperatures. ALMA comes into its own in the acquisition of redshift confirmation through spectroscopic line observations of individual targets. Unfortunately, ALMA's high angular resolution is also accompanied by a small field of view, limiting its capacity for large surveys.

Some planned missions  will be dedicated to targeting the peak of the warm dust SEDs at moderate redshift.
These include CCAT-prime/FYST \citep[][]{2023ApJS..264....7C}, AtLAST \citep{2020SPIE11445E..2FK}, Antartic THz telescope \citep{2022MNRAS.509.2258C} and the Origins Space Telescope \citep{2020arXiv201202731W}. However, until and unless spectroscopic redshifts, or photometric samples with far lower contamination, are available, the interpretation of the dust-temperature redshift curve will always be called into question.

\section*{Acknowledgements}
ERS and GTJ acknowledge support from the UK Science and Technology Facilities Council (STFC) through consolidated grant ST/T000406/1 and a doctoral studentship, respectively.
This work made use of the University of Warwick Scientific Computing Research Technology Platform (SCRTP) and Astropy\footnote{\url{https://www.astropy.org/}}, a community-developed core Python package for Astronomy \citep{astropy:2013,astropy:2018}.

\section*{Data Availability}
All photometric and modelling data underlying these results are publically available. Specific results of fits will be made available via the BPASS website at \url{www.warwick.ac.uk/bpass} or \url{bpass.auckland.ac.nz}, or
 by reasonable request to the first author.


\bibliographystyle{mnras}
\bibliography{references} 

\bsp	
\label{lastpage}




\clearpage
\setcounter{page}{0}
    \pagenumbering{roman}
    \setcounter{page}{1}

\appendix

\section{Deriving Photometric Redshifts} \label{sec:photo_append}
\begin{figure}
    \includegraphics[width=\columnwidth]{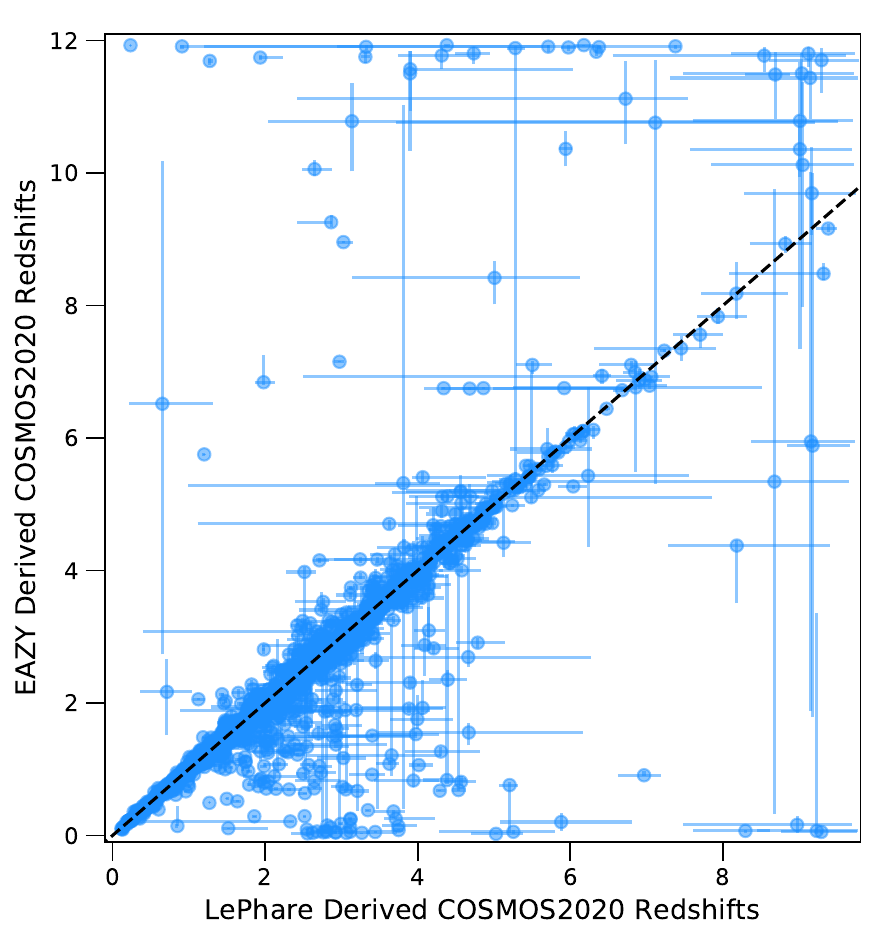}
    \caption{Derived redshifts taken from the COSMOS2020 catalogue using the {\sc{LePhare}} and EAZY tools for all galaxies remaining in the sample after the chance alignment probability cut. The black dashed line indicates when the two values are in agreement.}
    \label{fig:cosmos_z_comp}
\end{figure}

The COSMOS2020 catalogue derives photometric redshifts using two methods; {\sc{LePhare}} profile-fitting tool \citep{2002MNRAS.329..355A,2006A&A...457..841I} and the EAZY photometric redshift code \citep{2008ApJ...686.1503B}. We use the {\sc{LePhare}} redshifts in our main analysis. 
We compare the  photometric redshift derived by the COSMOS2020 team using each method in Fig.~\ref{fig:cosmos_z_comp}.
Whilst most of the derived redshifts are in reasonable agreement, catastrophic failures are evident, as was seen when comparing COSMOS2020 against the \bagpipes-derived redshifts in Fig.~\ref{fig:z_comp}. This demonstrates the fundamental limitations of the photometric redshift technique, in which small differences in the presumed template set, fitting methodology and prior assumptions can change the prioritisation of peaks in the posterior probability distribution.

Catastrophic failures of photometric redshift estimation can work in either direction. In a number of cases, {\sc{LePhare}} favours a high redshift solution while EAZY gave a low redshift solution, and vice versa. These catastrophic failures are not well captured by formal statistical uncertainties.

High-redshifts derived photometrically are particularly problematic as low-$z$ galaxy templates can sometimes mimic the same photometric observational properties, for example when a strong rest-optical emission line or  spectral break is misinterpreted as an ultraviolet feature at higher redshift.
In these cases, and in the absence of further information, the low redshift solution is to be preferred due to the strong luminosity bias against high redshift galaxies, and the equally strong evolution in the typical number density of sources at a given mass. Spectroscopic confirmation of complete samples of  galaxies is the most reliable method to determine redshifts, however comprehensive spectroscopic coverage is not plausible for deep field, distant galaxy surveys with current technology.

\begin{figure}
    \includegraphics[width=\columnwidth]{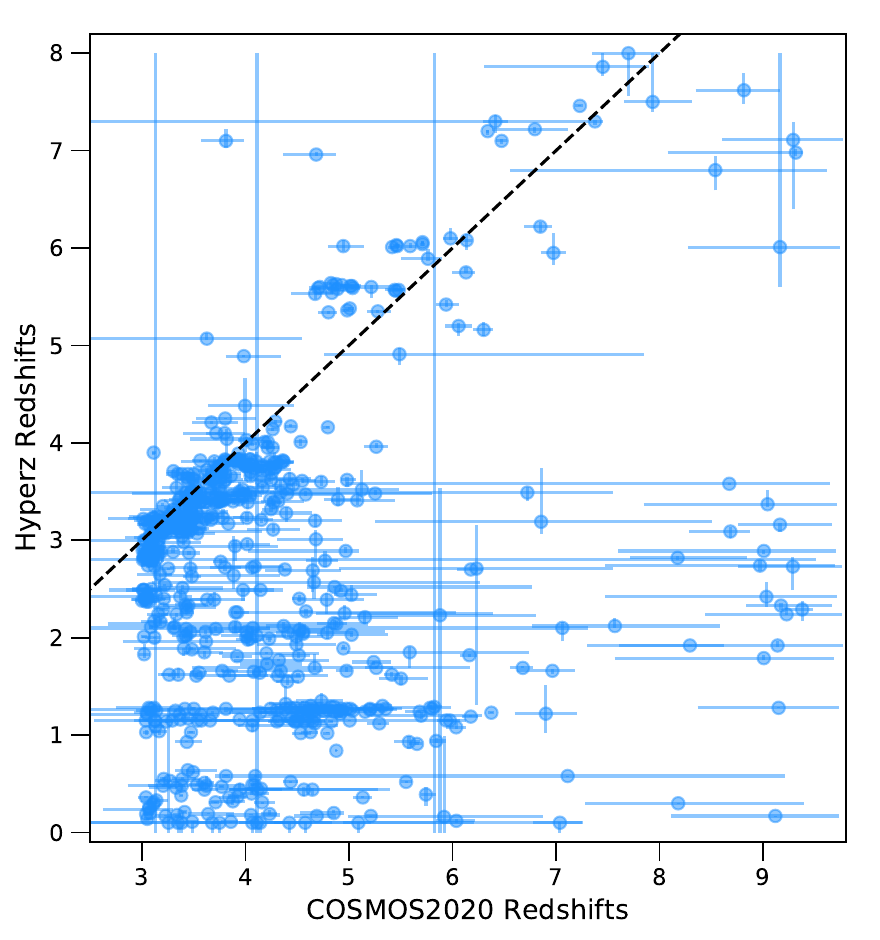}
    \caption{Derived redshifts using the Hyperz package compared to those from the COSMOS2020 catalogue, derived using the {\sc{LePhare}} profile-fitting tool. The black dashed line indicates when the two values are in agreement. While all galaxies are fit, only those with with COSMOS2020 derived redshifts of $z>3$ are shown.}
    \label{fig:hyp_z_comp}
\end{figure}

To provide an additional, independent cross check. We also derive a set of photometric redshifts for the same objects using the Hyperz package \citep{2000A&A...363..476B}. This relatively simple package only permits 20 UV-optical-NIR photometric bands per object, so broadband magnitudes and a limited set of narrow bands were selected. We compare the new photometric redshifts to the COSMOS2020 {\sc{LePhare}} redshifts in Fig.~\ref{fig:hyp_z_comp}.
This comparison has a larger number of  catastrophic failures, with most of them having Hyperz-derived lower photometric redshift estimates than the COSMOS2020 ones, demonstrating the existence of a plausible secondary solution in the low redshift regime.
Ultimately, this analysis highlights how uncertain photometric redshifts can be, and how the choice of fitting tool and the selected priors can have a direct result on the inferred redshifts.

\end{document}